\documentclass[onecolumn,draftcls,12pt]{IEEEtran}
\usepackage{times}
\usepackage{amssymb,amsmath,amsthm}
\usepackage{amsmath,epsfig,graphics,subfigure}
\usepackage{algorithm,algorithmic}
\usepackage{cite}

\begin{document}
\title{Joint Channel Estimation and Data Detection for Multihop OFDM Relaying System under Unknown Channel Orders and Doppler Frequencies}

\author{Rui Min and Yik-Chung Wu$^*$
\thanks{Rui Min and Yik-Chung Wu are with the Department of Electrical and Electronic Engineering, The University of Hong Kong, Pokfulam Road, Hong Kong. Email:\{minrui,ycwu\}@eee.hku.hk.}
\thanks{$^*$The corresponding author is Yik-Chung Wu.}}
\maketitle

\begin{abstract}
In this paper,  channel estimation and data detection for multihop relaying orthogonal frequency division multiplexing (OFDM) system is investigated under time-varying channel.  Different from previous works, which highly depend on the statistical information of the doubly-selective channel (DSC) and noise to deliver accurate channel estimation and data detection results, we focus on more practical scenarios with unknown channel orders and Doppler frequencies.
Firstly, we integrate the multilink, multihop channel matrices into one composite channel matrix. Then, we formulate the unknown channel using generalized complex exponential basis expansion model (GCE-BEM) with a large oversampling factor to introduce channel sparsity on delay-Doppler domain. To enable the identification of nonzero entries, sparsity enhancing Gaussian distributions with Gamma hyperpriors are adopted. An iterative algorithm is developed under variational inference (VI) framework. The proposed algorithm iteratively estimate the channel, recover the unknown data using Viterbi algorithm and learn the channel and noise statistical information, using only limited number of pilot subcarrier in one OFDM symbol.
Simulation results show that, without any statistical information, the performance of the proposed algorithm is very close to that of the optimal channel estimation and data detection algorithm, which requires specific information on system structure, channel tap positions, channel lengths, Doppler shifts as well as noise powers.
\end{abstract}
\begin{IEEEkeywords}
Doubly-selective channel, Channel estimation, Data detection, Variational inference, Orthogonal frequency division multiplexing, Multihop relaying system
\end{IEEEkeywords}

\section{Introduction}\label{Intro}
Next generation broadband system aims to support higher levels of mobility, connectivity and efficiency. Multihop relaying system is a perfect suit for such requirement due to their benefits in easy deployment, enhanced connectivity, flexible adaptability, and increased capacity. On the other hand, orthogonal frequency division multiplexing (OFDM) has been adopted as the transmission scheme for many next generation broadband standards, such as WiMAX, LTE and IEEE 802.16. These result in the need to develop receiver algorithms for multihop OFDM system under high mobility. With high mobility, the broadband wireless channel is both frequency-selective and time-varying, a.k.a. doubly-selective. The channel responses vary sample by sample, which destroy the orthogonal property among subcarriers and causes intercarrier interference (ICI). Besides, the relaying system structure and channel statistical information are generally unknown to the receiver, due to flexible configuration of relaying paths. These poses strong challenges to channel estimation and data detection of OFDM relaying system under high mobility.

Over doubly-selective channel (DSC), channel estimation and data detection for single-hop OFDM systems has been considered in \cite{highmo2,highmo3,highmo4,highmo5,ALLBEM,ModelReduction}, where the two tasks are treated separately.
In \cite{highmo2,highmo3,highmo4,highmo5}, the frequency domain channel matrix is approximated with a diagonal matrix under the assumption of small normalized Doppler frequency. The resulting algorithm would produce poor channel estimate and subsequently degrade the data detection performance for fast time-varying DSC. In view of that, \cite{ALLBEM} and \cite{ModelReduction} assumed a banded frequency-domain channel matrix, thus achieve a better channel modeling accuracy. However, due to the ICI introduced in frequency domain, pilots and data would interfere each other. This leads to the interdependence between channel estimation and data detection, and joint processing of them is necessary.

Research on multihop channel estimation is still limited especially for DSC, due to the complicated system structure and uncertain time-varying channel property. Among the limited existing works, \cite{Relay1} studied the two-way relaying (TWR) system
under frequency-nonselective time-varying channel, and complex exponential basis expansion model (CE-BEM) was used to reduce the number of channel parameters. However, \cite{Relay1} only considered single carrier system and extension to multicarrier system is not straightforward due to additional ICI. More recently, in \cite{Helen3}, an
iterative algorithm for data detection and channel estimation was proposed for dual-hop
amplify-and-forward (AF) OFDM system. With complete information
of channel and noise in each hop, the data detection results were
very close to the ideal case. Unfortunately, this work did not consider multihop relaying system, which has a more complicated structure. More importantly, it requires the destination receiver to have full statistical information of all channels and
noise powers, which might not be readily available in practice.

In this paper, we study the channel estimation and data detection of OFDM-based multihop  AF relaying system under high mobility, with special focus on unknown channel orders and Doppler frequencies. Based on the fact that the combined channel information is sufficient for data detection, and in order to reduce computation load on relays and time delay of the whole system, no channel estimation is performed at the relays. Different from previous works which highly rely on information of system structure, channel tap positions, channel lengths and Doppler frequencies of all channels, as well as noise powers at all receivers, we propose to solve the problems with none of the above information. By first expanding the composite source-relay-destination channel using generalized complex exponential basis expansion model (GCE-BEM) with a large oversampling factor, we introduce channel sparsity on delay-Doppler domain. Then sparsity enhancing Gaussian distributions with Gamma hyperpriors are adopted for channel estimation to enable the identification of nonzero elements. An iterative algorithm is proposed based on variational inference (VI) framework to iteratively  estimate the channel, recover the unknown data using Viterbi algorithm and learn the channel and noise statistical information, using only limited number of pilot subcarrier in one OFDM symbol. Simulation results show that the performance of the proposed algorithm is very close to that of an optimal algorithm, which requires detailed statistical information on channels and noises.

The rest of the paper is organized as follows. The OFDM-based multihop relaying system is introduced in Section \ref{section1}. Then in Section \ref{GCEreform}, the channel matrices of all the hops are integrated into one concise composite channel matrix and reformulated using GCE-BEM with a large oversampling factor. The iterative channel estimation and data detection algorithm is developed under VI framework in Section \ref{sbl}. And in Section \ref{initial}, least squares (LS) channel estimator and equalizer are derived to obtain the initial parameters for the proposed iterative algorithm. Simulation results of the proposed algorithm are presented in Section \ref{simulation}. Finally, this paper is concluded in Section \ref{conclusion}.

\emph{Notations}:
Boldface uppercase and lowercase letters will be used for matrices and vectors, respectively.
Superscripts $H$, $T$ and $*$ denote Hermitian, transpose and conjugate, respectively.
The symbol $\mathbf{I}_N$ represents the $N\times N$ identity matrix.
Symbol $\mathbf{e}_l$ denotes the  vector with structure given as
$\left[ \mathbf{0}_{1\times l}, 1, \mathbf{0}_{1\times (N-l-1)}\right]^T$, where $\mathbf{0}_{1\times l}$ is the $l$ dimension all-zero row vector.
$\mathrm{diag}\{\mathbf{x}\}$ stands for the diagonal matrix with vector $\mathbf{x}$
on its diagonal.
The notation $[\mathbf{X}]_{m_1:m_2,n_1:n_2}$ represents the submatrix of
$\mathbf{X}$ consists of entries on the $m_1$-to-$m_2^{th}$ rows and $n_1$-to-$n_2^{th}$ columns. $\mathbb{E}\{\cdot\}$ denotes the expectation while $\mathrm{Tr}\{\mathbf{X}\}$ and $\mathrm{det}\{\mathbf{X}\}$ are the trace and the determinant of the square matrix $\mathbf{X}$. $\mathrm{Re}\{\cdot\}$ denotes the real part. And $\lceil x\rceil$ rounds $x$ to the nearest integer greater than or equal to $x$. Finally, $\mathbf{F}$ represents the discrete Fourier transform (DFT) matrix with $[\mathbf{F}]_{m,n} = \frac{1}{\sqrt{N}}e^{-j2\pi mn/N}$.
\section{System Model}\label{section1}
\begin{figure}[!htb]
\begin{center}
  \includegraphics[width=0.6\textwidth]{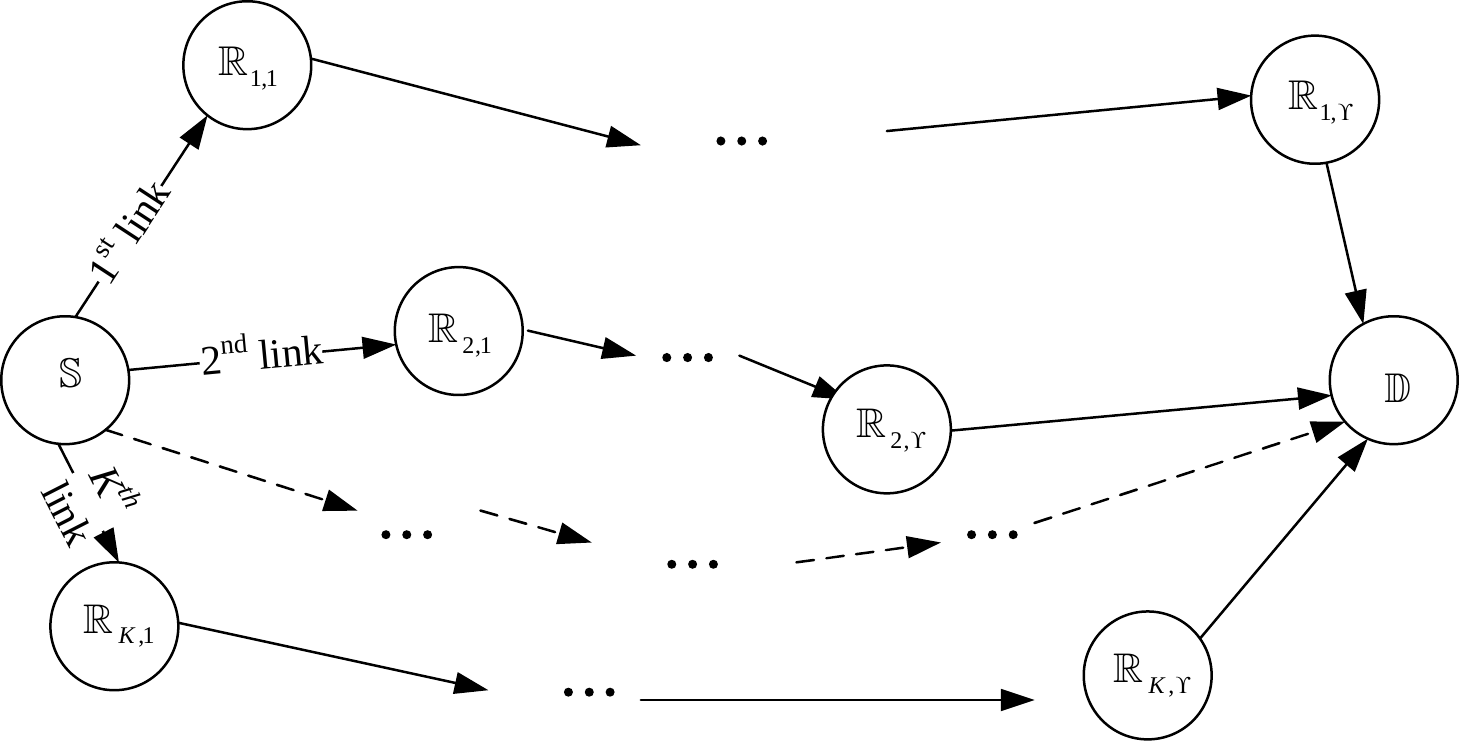}\\
  \caption{Multihop Cooperative Communication System}\label{MultihopModel}
\end{center}
\end{figure}

In this paper, we consider a multihop relaying system consists of a source $\mathbb{S}$, a destination $\mathbb{D}$ and a number of relays scattered in the middle. Each of them is equipped with single antenna. Without loss of generality, we assume the relays work cooperatively to form $K$ links, each of them consisting of $\Upsilon+1$ hops. Apart from the $K$ relaying paths, there is no other link between $\mathbb{S}$ and $\mathbb{D}$, and all relays employ the AF scheme. Denoting the relay on the ${k^{th}}$ link connecting the $\rho^{th}$ and the $(\rho+1)^{th}$ hop as $\mathbb{R}_{k,\rho}$, the relaying system is shown in Figure \ref{MultihopModel}.

The channel of each hop is assumed to be doubly-selective channel (DSC). Specifically, at the $\rho^{th}$ hop of the $k^{th}$ relaying path, the channel consists of $N_{k,\rho}$ independent nonzero channel taps with maximum delay of $(L_{\max}^{k,\rho}-1)T_s$, where $T_s$ is the sample interval. We consider the general situation that the channel taps are not necessarily consecutive, so that we have $N_{k,\rho}\leq L_{\max}^{k,\rho}$. Let $\bar h_{k,\rho}(n,l)$ be the $l^{th}$ tap of that channel at time $nT_s$. For a given $\rho$ and $k$, the channel taps are independent and each one being a zero-mean complex Gaussian process with bandlimited power spectral density within $[-f_{k,\rho}(l), f_{k,\rho}(l)]$, where $f_{k,\rho}(l)$ is the maximum Doppler shift of the $l^{th}$ tap. In general, $f_{k,\rho}(l)$ may be distinct for different $l$, since each tap results from signal transmission through a different physical scattering. Furthermore, it is assumed that the channels for different links $k$ and hops $\rho$ are independent from each other.
\subsection{OFDM Signal Transmitted from $\mathbb{S}$}
In an OFDM system, the frequency domain source data $\mathbf{x} = [x(0), \ldots, x(N-1)]^T$ is first transformed to the time domain data $\mathbf{s = F}^H\mathbf{x}$, where $\mathbf{F}$ represents the discrete Fourier transform (DFT) matrix. In order to facilitate channel estimation and data detection, pilots are inserted in the frequency domain as
\begin{equation}
  x(n) = \left\{
  \begin{array}{ccc}
    x_p(n) & \forall & n\in \mathfrak{I}_p\\
    x_d(n) & \forall & n\in \mathfrak{I}_d,
  \end{array}
  \right.
\end{equation}
where $\mathfrak{I}_d$ is the index set of the $N_d$ unknown data symbols, $\mathfrak{I}_p$ is the index set of the $N_p$ pilot symbols and we have $N = N_d+N_p$.
 In matrix form, $\mathbf{x}$ can be represented as
\begin{equation}\label{datastructure}
  \mathbf{x = E}_d\mathbf{x}_d + \mathbf{E}_p\mathbf{x}_p,
\end{equation}
where $\mathbf{E}_d$ and $\mathbf{E}_p$, with dimensions $N\times N_d$ and $N\times N_p$, respectively, map $\mathbf{x}_d$ and $\mathbf{x}_p$ to their corresponding subcarriers.
Before transmission, a Cyclic Prefix (CP) of length $L_{cp}$  is added at the beginning of $\mathbf{s}$ to prevent intersymbol interference (ISI). Since the OFDM signal goes through a number of relays before reaching the destination, $L_{cp}$ should be larger than the maximum channel length among all the relaying paths, denoted as $L_{\max}= \max\limits_ k(\sum_{\rho=1}^{\Upsilon+1}L^{k,\rho}_{\max}-\Upsilon)$.
\subsection{Received OFDM Signal}
In AF relaying system, each relay merely amplify the received signal before passing the signal to the next relay or destination.
For the ${k^{th}}$ relaying path, the signal received at $\mathbb{R}_{k,1}$ is given by
\begin{equation}
  r_{k,1}(n) = \sum_{l=0}^{L_{\max}^{k,1}-1}\bar h_{k,1}(n,l)s(n-l) + w_{k,1}(n),
\end{equation}
where $w_{k,1}(n)$ is the additive white Gaussian noise (AWGN) with power $\varpi_{k,1}^2$. Upon receiption, the relay amplifies the incoming signal as \cite{AF1}
\begin{equation}
  z_{k,1}(n) = \varsigma_{k,1}r_{k,1}(n)
\end{equation}
and then transmits $z_{k,1}(n)$ to  the next relay $\mathbb{R}_{k,2}$ through channel $\bar h_{k,2}(n,l)$.
The received signal is then represented as
\begin{equation}
  r_{k,2}(n) = \sum_{l=0}^{L_{\max}^{k,2}-1}\bar h_{k,2}(n,l)z_{k,1}(n-l) + w_{k,2}(n),
\end{equation} with $w_{k,2}(n)$ being the AWGN with power $\varpi_{k,2}^2$. The signal is then amplified as $z_{k,2}(n) = \varsigma_{k,2}r_{k,2}(n)$ and transmitted to $\mathbb{R}_{k,3}$ and so on.
Then, at the $\Upsilon^{th}$ relay, the amplified signal is transmitted to the destination. Finally at destination $\mathbb{D}$, the received signal is given by
\begin{equation}
  \tilde y(n) = \sum_{k=1}^{K}\sum_{l = 0}^{L_{\max}^{k,\Upsilon+1}-1}\bar h_{k,\Upsilon+1}(n,l)z_{k,\Upsilon}(n-l)+w_{d}(n),
\end{equation} where AWGN $w_{d}(n)$ has power $\varpi_d^2$.
Upon reception, the CP $[\tilde y(-L_{cp}), \ldots, \tilde  y(-1)]^T$ is removed and the received signal $\mathbf{\tilde y} = [\tilde y(0), \ldots, \tilde y(N-1)]^N$ can be written in matrix form as
\begin{equation}
  \mathbf{\tilde y}=\sum_{k=1}^K\mathbf{\bar H}_{k,\Upsilon+1}\mathbf{z}_{k,\Upsilon} + \mathbf{w}_d,
\end{equation} where $\mathbf{z}_{k,\Upsilon} = [z_{k,\Upsilon}(-(L_{\max}^{k,\Upsilon+1} -1)),\ldots, z_{k,\Upsilon}(0),\ldots, z_{k,\Upsilon}(N-1) ]^T$, $\mathbf{w}_d$ is the noise vector with elements $w_d(n)$, and $\mathbf{\bar H}_{k,\Upsilon+1}$ is an $N\times (N+L_{\max}^{k,\Upsilon+1}-1)$ channel matrix given by
\begin{equation}\label{hbargammaplus1}
  \mathbf{\bar H}_{k,\Upsilon+1} = \left[
  \begin{array}{l}
  \bar h_{k,\Upsilon+1}(0, L_{\max}^{k,\Upsilon+1}-1)\qquad\ldots \qquad   \bar h_{k,\Upsilon+1}(0,0) \qquad \\
  \quad   \bar h_{k,\Upsilon+1}(1, L_{\max}^{k,\Upsilon+1}-1)\qquad\ldots \qquad   \bar h_{k,\Upsilon+1}(1,0) \quad\\
  \qquad\quad\quad\quad\qquad\qquad\qquad\qquad\cdots\\
  \qquad\quad   \bar h_{k,\Upsilon+1}(N-1, L_{\max}^{k,\Upsilon+1}-1)\quad\ldots \quad   \bar h_{k,\Upsilon+1}(N-1,0)
  \end{array}
  \right].
\end{equation}
Furthermore, $\mathbf{z}_{k,\Upsilon}$ can be written in terms of $\mathbf{z}_{k,\Upsilon-1}$ as
\begin{equation}
  \mathbf{z}_{k,\Upsilon} =  \varsigma_{k,\Upsilon}\mathbf{\bar H}_{k,\Upsilon}\mathbf{z}_{k,\Upsilon-1} + \varsigma_{k,\Upsilon}\mathbf{w}_{k,\Upsilon},
\end{equation}
with $\mathbf{z}_{k,\Upsilon-1} = [z_{k,\Upsilon-1}(-(L_{\max}^{k,\Upsilon+1}+L_{\max}^{k,\Upsilon})+2), \ldots, z_{k,\Upsilon-1}(0), \ldots, z_{k,\Upsilon-1}(N-1)]^T$, $\mathbf{w}_{k,\Upsilon}$ is the corresponding noise vector, and $\mathbf{\bar H}_{k,\Upsilon}$ is an $(N+L_{\max}^{k,\Upsilon+1}-1)\times(N+L_{\max}^{k,\Upsilon+1}+L_{\max}^{k,\Upsilon}-2)$ matrix given by
\begin{equation}\label{hbargamma}
  \mathbf{\bar H}_{k,\Upsilon} = \left[
  \begin{array}{l}
  \bar h_{k,\Upsilon}(1-L_{\max}^{k,\Upsilon+1}, L_{\max}^{k,\Upsilon}-1)\qquad\ldots \qquad   \bar h_{k,\Upsilon}(1-L_{\max}^{k,\Upsilon+1},0) \\
  \quad   \bar h_{k,\Upsilon}(2-L_{\max}^{k,\Upsilon+1}, L_{\max}^{k,\Upsilon}-1)\qquad\ldots \qquad   \bar h_{k,\Upsilon}(2-L_{\max}^{k,\Upsilon+1},0) \\
  \qquad\qquad\quad\quad\quad\qquad\qquad\qquad\qquad\cdots\\
  \qquad\quad \qquad\qquad   \bar h_{k,\Upsilon}(N-1, L_{\max}^{k,\Upsilon}-1)\quad\ldots \quad   \bar h_{k,\Upsilon}(N-1,0)
  \end{array}
  \right].
\end{equation}
Tracing back to the $1^{st}$ hop, we have $\mathbf{z}_{k,1} = \varsigma_{k,1}\mathbf{\bar H}_{k,1}\mathbf{s}_k + \varsigma_{k,1}\mathbf{w}_{k,1}$,
where $\mathbf{\bar H}_{k,1}$ is an $(N+ \sum_{\rho=2}^{\Upsilon+1}L_{\max}^{k,\rho}-\Upsilon)\times(N+ \sum_{\rho=1}^{\Upsilon+1}L_{\max}^{k,\rho}-\Upsilon-1)$ channel matrix with structure the same as (\ref{hbargammaplus1}) and (\ref{hbargamma}),
and $\mathbf{s}_k = \mathbf{E}_k\mathbf{s}$ with $\mathbf{E}_k = [[\mathbf{I}_N]_{1:N,(N- \sum_{\rho=1}^{\Upsilon+1}L_{\max}^{k,\rho}+\Upsilon+2):N}, \mathbf{I}_N]^T$ characterizing the effect of the CP.
Based on the above derivations, the received signal vector $\mathbf{\tilde y}$ is
\begin{eqnarray}\label{system}
  \mathbf{\tilde y} &=& \underbrace{\sum_{k=1}^K \left[\left(\prod_{\rho=1}^{\Upsilon} \varsigma_{k,\rho}\right) \left( \mathbf{\bar H}_{k,\Upsilon+1}\ldots \mathbf{\bar H}_{k,1}\right) \mathbf{E}_k \right]}_{\triangleq\mathbf{ H}}\mathbf{F}^H\mathbf{x}\nonumber \\
   &&+\underbrace{\sum_{k=1}^K \left[\sum_{\rho=1}^{\Upsilon}\left(\left(\prod_{\varrho =\rho}^
   {\Upsilon}\varsigma_{k,\varrho}\right)\left(\mathbf{\bar H}_{k,\Upsilon+1}\ldots\mathbf{\bar H}_{k,\rho+1}\right)
   \right)\mathbf{w}_{k,\rho}\right] +\mathbf{w}_d}_{\triangleq\mathbf{\tilde v}},
\end{eqnarray}
where $\mathbf{H}$ represents the composite channel matrix and $\mathbf{\tilde v}$ represents the composite noise effect.

\section{Reformulation of the Composite Channel Matrix}\label{GCEreform}
In order to estimate the channel and detect the data, it is important to investigate the structure of the channel matrix $\mathbf{H}$. Writing $\mathbf{H} = \sum_{k = 1}^K \left(\prod_{\rho=1}^{\Upsilon} \varsigma_{k,\rho}\right) \mathbf{H}_k$, where $\mathbf{H}_k = \mathbf{\bar H}_{k,\Upsilon+1}\mathbf{\bar H}_{k,\Upsilon}\ldots\mathbf{\bar H}_{k,1}\mathbf{E}_k$.
To find out the structure of $\mathbf{\bar H}_{k,\Upsilon+1}\mathbf{\bar H}_{k,\Upsilon}\ldots \mathbf{\bar H}_{k,1}\mathbf{E}_k$, we start from $\mathbf{\bar H}_{k,\Upsilon+1}$ and $\mathbf{\bar H}_{k,\Upsilon}$ with their expressions given in (\ref{hbargammaplus1}) and (\ref{hbargamma}), respectively. Each matrix represents the linear convolution of a time-varying channel and the matrix multiplication expresses the convolution effect of two time-varying channels. Therefore the resulting matrix $\mathbf{\bar H}_{k,\Upsilon+1}\mathbf{\bar H}_{k,\Upsilon}$ will also be in the form of (\ref{hbargammaplus1}) and (\ref{hbargamma}), except that the resulting channel length of the new time-varying channel is now being $L_{\max}^{k,\Upsilon+1}+L_{\max}^{k,\Upsilon}-1$.

Similarly, multiplying $\mathbf{\bar H}_{k,\Upsilon-1}$ to $\mathbf{\bar H}_{k,\Upsilon+1}\mathbf{\bar H}_{k,\Upsilon}$ from the right, the result $\mathbf{\bar H}_{k,\Upsilon+1}\mathbf{\bar H}_{k,\Upsilon}\mathbf{\bar H}_{k,\Upsilon-1}$ will be an $N\times (N+\sum_{\rho = \Upsilon-1}^{\Upsilon+1}-3)$ matrix, with equivalent channel length of $\sum_{\rho = \Upsilon-1}^{\Upsilon+1}L_{\max}^{k,\rho}-2$, due to the convolution effect.
Continuing the matrix multiplication, we have $\mathbf{\bar H}_{k,\Upsilon+1}\ldots\mathbf{\bar H}_{k,1}$ being an $N\times (N+\sum_{\rho=1}^{\Upsilon+1}L_{\max}^{k,\rho}-\Upsilon-1)$ matrix, with equivalent channel length of $\sum_{\rho=1}^{\Upsilon+1}L_{\max}^{k,\rho}-\Upsilon$.
And eventually $\mathbf{E}_k$ moves the $\sum_{\rho = 1}^{\Upsilon+1}L_{\max}^{k,\rho}-\Upsilon-1$ columns from the left part of $\mathbf{\bar H}_{k,\Upsilon+1}\ldots\mathbf{\bar H}_{k,1}$ to the upper right corner. The resulted composite channel matrix ${\mathbf{H}}_{k}$ is an $N\times N$ circular convolution matrix of a time-varying channel with equivalent channel length of $\sum_{\rho=1}^{\Upsilon+1}L_{\max}^{k,\rho}-\Upsilon$.
Thus $\mathbf{H}$, as the weighted sum of $\mathbf{H}_k$'s, has the same circular convolution matrix structure of a time-varying channel with length $L_{\max}= \max\limits_ k(\sum_{\rho=1}^{\Upsilon+1}L^{k,\rho}_{\max}-\Upsilon)$:
\begin{equation}
  \mathbf{H} = \left[
  \begin{array}{ccccc}
    \bar \mu(0, 0)& \mathbf{0} & \bar \mu(0,L_{\max}-1) & \ldots & \bar \mu(0,1)\\
    \bar \mu(1,1)&\bar \mu(1,0)&\mathbf{0}&\bar\mu (1,L_{\max}-1)&\ldots\\
    &\ldots&\ldots&\ldots&\\
    &\mathbf{0}&\bar\mu(N-1, L_{\max}-1)&\ldots&\bar\mu(N-1, 0)
  \end{array}
  \right],
\end{equation}
or equivalently \begin{equation}
  \mathbf{H} = \sum_{l = 0}^{L_{\max}-1}\mathrm{diag}\{\bar{\boldsymbol\mu}_l\}\mathbf{P}(l),
\end{equation}
where $\bar{\boldsymbol\mu}_l = [\bar\mu(0,l),\ldots,\bar\mu(N-1,l)]^T$ consists of all the composite channel coefficients of the $l^{th}$ tap and $\mathbf{P}(l) = [\mathbf{e}_l, \ldots, \mathbf{e}_{N-1}, \mathbf{e}_0,\ldots, \mathbf{e}_{l-1}]$. Thus (\ref{system}) becomes
\begin{equation}
\mathbf{  \tilde y} = \sum_{l = 0}^{L_{\max}-1}\mathrm{diag}\{\bar{\boldsymbol\mu}_l\}\mathbf{P}(l) \mathbf{F}^H\mathbf{x+\tilde v}.\label{sm4klem}
\end{equation}

It should be noticed that, the receiver knows neither the individual channel  information of each hop nor the statistical information about the composite channel. This is a natural assumption, as the channels are time-varying and depend on the speed of transceivers and the environment around them. In particular, the receiver has no knowledge on the composite channel tap positions (if the tap positions of individual channel are not consecutive) and the maximum Doppler shift $f_{\max} = \max\limits_k\big(\sum_{\rho = 1}^{\Upsilon+1}\max\limits_{l\in[0,L^{k,\rho}_{\max}]}f_{k,\rho}(l)\big)$. Furthermore, the noise power at each relay $\mathbb{R}_{k,\rho}$ is not available to the receiver either. As a result, the receiver has no information on the composite noise power.

In order to proceed, we propose to calculate an upper bound on the maximum Doppler shift and the delay for the composite channel. Let $v_{\max}$ be the maximum relative velocity between two units in any hop in the relaying system. Since $v_{\max}f_c/ c\geq f_{k,\rho}(l)$ for all $k$, $\rho$ and $l$, where $f_c$ and $c$ are the carrier frequency and the speed of light, respectively,
we have $f_{\max}\leq f_{U}=(\Upsilon+1)v_{\max}f_c/c$. And in the delay domain, the best the receiver knows is that $L_{cp}$ is chosen large enough to avoid ISI. Thus $L_{\max}\leq L_{cp}$ and all the nonzero taps fall in the range of $\{0, \ldots, L_{cp}-1\}$.
With the ranges of the delay-Doppler domain defined for the composite channel, we can expand the channel with generalized complex exponential basis expansion model (GCE-BEM) as follows
\begin{equation}\label{mhbem}
  \bar \mu(n,l) = \sum_{q=-Q}^{Q} \mu_q(l)e^{j2\pi qn/VN},\quad l = 0,\ldots,L_{cp}-1,\quad n = 0, \ldots, N-1,
\end{equation}
where $Q = \lceil VNf_UT_s\rceil$ and $V$ is the oversampling factor, and $\mu_q(l)$ is the GCE-BEM coefficient.
It should be noticed that $\mu_q(l)=0$ in two conditions: 1) $\bar\mu(n,l)=0$; 2) $|q|>VN f_{\max}T_s$. 

From (\ref{mhbem}), the vector $\bar{\boldsymbol\mu}_l$ can be expressed as
$\bar{\boldsymbol\mu}_l=\sum_{q =-Q}^{Q}\boldsymbol\varphi(q)\mu_q(l)$,
where $\boldsymbol\varphi(q) = [1, e^{j2\pi q/VN}$ $,\ldots, e^{j2\pi q(N-1)/VN}]^T$ denotes the $q^{th}$ basis vector.
Putting this result into (\ref{sm4klem}), taking the DFT on the signal $\mathbf{\tilde y}$ and replacing the unknown $L_{\max}$ with $L_{cp}$, we have
\begin{eqnarray}\label{ybem1}
  \mathbf{y} = \mathbf{F\tilde y} &=&
\sum_{l=0}^{L_{cp}-1}\mathbf{F}\mathrm{diag}\{\sum_{q =-Q}^{Q}\boldsymbol\varphi(q)\mu_q(l)\}\mathbf{P}(l)\mathbf{F}^H\mathbf{x+v}\nonumber\\
&=&\sum_{l=0}^{L_{cp}-1}\sum_{q =-Q}^{Q}[\mathbf{F}\mathrm{diag}\{\boldsymbol\varphi(q)\}\mathbf{P}(l)\mathbf{F}^H\mathbf{x}]\mu_q(l)+\mathbf{v},
\end{eqnarray}
where $\mathbf{v=F\tilde v}$ represents the noise vector after DFT.
Let $\boldsymbol\mu_q = [\mu_q(0), \ldots, \mu_q(L_{cp}-1)]^T$, then (\ref{ybem1}) can be written as
\begin{eqnarray}
  \mathbf{y} = \sum_{q =-Q}^{Q} [\underbrace{\mathbf{F}\mathrm{diag}\{\boldsymbol\varphi(q)\}\mathbf{P}(0)\mathbf{F}^H\mathbf{x},\ldots, \mathbf{F}\mathrm{diag}\{\boldsymbol\varphi(q)\}\mathbf{P}(L_{cp}-1)\mathbf{F}^H\mathbf{x}}_{\triangleq\mathbf{G}_q[\mathbf{x}]}]\boldsymbol\mu_q+\mathbf{v}.
\end{eqnarray}
Further define $\boldsymbol\mu = [\boldsymbol\mu_{-Q}^T, \ldots, \boldsymbol\mu_Q^T]^T$ and let $\mathbf{G[x]}=[\mathbf{G}_{-Q}[\mathbf{x}],\ldots,\mathbf{G}_{Q}[\mathbf{x}]]$, thus we have
\begin{equation}\label{SM1}
  \mathbf{y=G[x]}\boldsymbol\mu + \mathbf{v}.
\end{equation} On the other hand, from (\ref{system}), let $\mathbf{D[\boldsymbol\mu]=FHF}^H$, the system model can also be written as
\begin{equation}\label{SM2}
  \mathbf{y=D}[\boldsymbol\mu]\mathbf{x+v}.
\end{equation}
It is clear that $\mathbf{D}[\boldsymbol\mu]\mathbf{x} = \mathbf{G[x]}\boldsymbol\mu$.

\section{Iterative Channel Estimation and Data Detection}\label{sbl}
From the system model (\ref{SM1}) and (\ref{SM2}), the problem is to jointly estimate the composite channel BEM coefficients $\boldsymbol\mu$ and the unknown data $\mathbf{x}_d$, without the knowledge of the composite noise variance, denoted by $\varpi_v^2$. Since we have expanded the composite channel over an extended range in the delay-Doppler plane, we also want to make use of the prior information that most of the BEM coefficients will be zero (i.e., $\boldsymbol\mu$ is sparse). It is noticed from (\ref{SM1}) and (\ref{SM2}) that, estimation of channel requires knowledge of data and vice versa, thus leads to challenges in joint channel estimation and data detection. In this paper, a variational framework is adopted to iteratively improve the channel estimation and data detection results. Compared with other iterative frameworks, e.g., expectation-maximization (EM), VI is more general as it works within a complete Bayesian paradigm and gives a posterior distribution over all the parameters. Below, we first assign prior distributions to the unknown parameters.
\subsection{Prior Distributions of the Unknown Parameters}
First, the prior distribution of ${\boldsymbol\mu}$ is assumed to be Gaussian 
\begin{equation}\label{hprior}
  p({\boldsymbol\mu|\boldsymbol\alpha}) = \frac{1}{\pi^M\det (\mathbf{A}^{-1})}\exp \{{\boldsymbol\mu}^H\mathbf{A}{\boldsymbol\mu}\},
\end{equation}
where $M=(2Q+1)L_{cp}$, $\mathbf{A} = \mathrm{diag}\{\boldsymbol\alpha\}$ and $\boldsymbol\alpha = [\alpha_1, \ldots,\alpha_M]^T$ is a vector containing the inverse variance of the elements of $\boldsymbol\mu$.
Then a hyperprior for $\boldsymbol\alpha$ is specified as \cite{VRVM}
\begin{equation}\label{prioralpha}
  p( \alpha_j) = \mathrm{Ga}( \alpha_j|a_j,b_j)=b_j^{a_j} \alpha_j^{a_j-1}\exp(-b_j \alpha_j)/\Gamma(a_j),
\end{equation} with parameters $a_j,b_j$.
Although the Gaussian prior given by (\ref{hprior}) does not have strong probability peaks for sparsity promotion, by working with (\ref{prioralpha}), the marginal prior
 $p({\boldsymbol\mu})$ obtained by integrating out $\boldsymbol\alpha$ is a $t-$distribution,
 which nicely approximates a Laplace distribution \cite{VRVM}. Laplace
 prior is widely adopted in $L$1-norm regularization schemes like Basis Pursuit (BP) \cite{BP}.
 Unfortunately, using the Laplace prior directly does not lead to a tractable
 variational treatment \cite{VRVM}. As a result, BP is usually used in one-shot
 sparse channel estimation \cite{sparse1,sparse2,sparse3}. Furthermore, BP or BP
 denoising methods rely on the noise power information \cite{SBL1}, which is not
 known in our case. Thus, the above hierarchical prior structure, which
 is both sparsity promoting and analytically tractable, is a suitable alternative for
 our problem.

For $\mathbf{x}_d$, since we do not have knowledge on its value before observing the received signal, we set equal preference to all constellation points. Furthermore, due to the independent property among data elements, we have
\begin{equation}\label{pxd}
  p(\mathbf{x}_d) = \frac{1}{\mathcal{M}_d^{N_d}}\prod_{n =1}^{N_d}\Big[ \sum_{{\bar x}_d(n)\in \mathbb{C}_d}\delta({x}_d(n)-{\bar x}_d(n))\Big],
\end{equation}
where $\mathbb{C}_d$ is the constellation points of the modulation and $\mathcal{M}_d$ is the modulation order, e.g., $\mathcal{M}_d=4$ for QPSK.

Besides, the unknown noise power is assumed to obey a Gamma prior, such that it
can be learned under the variational framework. For ease of expression,
let $\beta = 1/\varpi_v^2$ and then
\begin{equation}\label{priorbeta}
  p(\beta) = \mathrm{Ga}(\beta|c,d) = d^c \beta^{c-1} \exp(-d\beta)/\Gamma(c),
\end{equation}
where $c,d$ are the parameters of the Gamma distribution.
In the absence of prior information, small values for hyperparameters are chosen, i.e., $a_j=b_j=c=d = 10^{-6}$, so as to produce uninformative priors for
 the channel and noise power \cite{VRVM}.
\subsection{Variational Inference}
With the introduced prior and hyperprior distributions, our aim is to jointly estimate ${\boldsymbol\mu},\boldsymbol
\alpha,\beta$ and $\mathbf{x}_d$. In Bayesian framework, this corresponds to maximizing
the posterior probability density function (pdf)
$p({\boldsymbol\mu},\boldsymbol\alpha, \beta,\mathbf{x}_d|\mathbf{y})$. However, this
pdf is in general very hard to be obtained in closed-form and the maximization of it is
inconvenient. In the VI framework,
a $Q({\boldsymbol\mu},\boldsymbol\alpha,\beta,\mathbf{x}_d)$ function, which is in
tractable form but closely represents $p({\boldsymbol\mu},\boldsymbol\alpha,\beta,
\mathbf{x}_d|\mathbf{y})$, is adopted to efficiently derive the estimation algorithm.
The optimal $Q({\boldsymbol\mu},\boldsymbol\alpha,\beta,\mathbf{x}_d)$ function can be
obtained by minimizing the \emph{free energy} function defined as \cite{FEF}:
\begin{equation}\label{freeenergy}
   \mathbb{F} = \int_{{\boldsymbol\mu},\boldsymbol\alpha,\beta,\mathbf{x}_d}
   Q({\boldsymbol\mu},\boldsymbol\alpha,\beta,\mathbf{x}_d)\log \frac{Q({\boldsymbol\mu},\boldsymbol\alpha,\beta,\mathbf{x}_d)}
   {p({\boldsymbol\mu},\boldsymbol\alpha,\beta,\mathbf{x}_d,\mathbf{y})}d{\boldsymbol\mu}d\boldsymbol\alpha d\beta d\mathbf{x}_d.
\end{equation}
Notice that, $p({\boldsymbol\mu},\boldsymbol\alpha,\beta,\mathbf{x}_d,\mathbf{y})$ is used instead of $p({\boldsymbol\mu},\boldsymbol\alpha,\beta,\mathbf{x}_d|\mathbf{y})$ because they are proportional and thus equivalent in free energy formulation.
According to the mean-field approximation \cite{meanfield}, $Q({\boldsymbol\mu},
\boldsymbol\alpha,\beta,\mathbf{x}_d)$ can be factorized into a product form, i.e., $Q({\boldsymbol\mu} ,\boldsymbol\alpha,\beta,\mathbf{x}_d) = Q({\boldsymbol\mu})Q(\boldsymbol\alpha)Q(\beta)Q(\mathbf{x}_d)$. This is equivalent to assuming that ${\boldsymbol\mu}, \boldsymbol\alpha, \beta$ and $\mathbf{x}_d$ are independent conditioned on $\mathbf{y}$ and will greatly simplify the iterative algorithm.
With the mean-field approximation, the variational free energy in (\ref{freeenergy}) is given by
\begin{eqnarray}
    \mathbb{F} &=&  \int_{{\boldsymbol\mu},\boldsymbol\alpha,\beta,\mathbf{x}_d}
    Q({\boldsymbol\mu},\boldsymbol\alpha,\beta,\mathbf{x}_d)\log \frac{Q({\boldsymbol\mu},\boldsymbol\alpha,\beta,\mathbf{x}_d)}
    {p(\mathbf{y}|{\boldsymbol\mu},\beta,\mathbf{x}_d)
    p({\boldsymbol\mu}|\boldsymbol\alpha)
    p(\boldsymbol\alpha)p(\beta)p(\mathbf{x}_d)}d{\boldsymbol\mu}d\boldsymbol\alpha d\beta d\mathbf{x}_d\nonumber\\
    &=& \int_{{\boldsymbol\mu}} Q({\boldsymbol\mu}) \log Q({\boldsymbol\mu})d{\boldsymbol\mu} + \int_{\boldsymbol\alpha}Q(\boldsymbol\alpha)\log
    Q(\boldsymbol\alpha)d\boldsymbol\alpha+\int_{\beta} Q(\beta)\log Q(\beta)d\beta\nonumber\\
    &&  + \int_{\mathbf{x}_d}Q(\mathbf{x}_d)\log Q(\mathbf{x}_d)d\mathbf{x}_d-\int_{{\boldsymbol\mu},\boldsymbol\alpha} Q({\boldsymbol\mu})Q(\boldsymbol\alpha)\log p({\boldsymbol\mu}|\boldsymbol\alpha)d{\boldsymbol\mu}d\boldsymbol\alpha
    -\int_{\boldsymbol\alpha} Q(\boldsymbol\alpha)\log p(\boldsymbol\alpha)d\boldsymbol\alpha\nonumber\\
    && -\int_{\beta}Q(\beta)\log p(\beta)d\beta - \int_{\mathbf{x}_d}Q(\mathbf{x}_d)\log p(\mathbf{x}_d)d\mathbf{x}_d\nonumber\\
    && -\int_{{\boldsymbol\mu},\beta,\mathbf{x}_d} Q({\boldsymbol\mu})Q(\beta)Q(\mathbf{x}_d) \log p(\mathbf{y}|{\boldsymbol\mu},\beta,\mathbf{x}_d)
    d{\boldsymbol\mu}d\beta d\mathbf{x}_d.\label{fe}
\end{eqnarray}

In order to calculate the free energy function given in (\ref{fe}), the likelihood
function $p(\mathbf{y}|\boldsymbol\mu,\beta,\mathbf{x}_d)$ and the form of $Q$ functions are needed,
in addition to (\ref{hprior}), (\ref{prioralpha}), (\ref{pxd}) and 
(\ref{priorbeta}). Since the noise is assumed to be AWGN, the likelihood
function is given by
\begin{equation}\label{likelihood}
  p(\mathbf{y}|{\boldsymbol\mu},\beta,\mathbf{x}_d)  = \left(\frac{\beta}{\pi}\right)^N\exp\{-\beta(\mathbf{y-G[x]}\boldsymbol\mu)^H(\mathbf{y-G[x]}\boldsymbol\mu)\}.
\end{equation}

For $Q(\mathbf{h})$, $Q(\boldsymbol\alpha)$, $Q(\beta)$ and $Q(\mathbf{x}_d)$, they represent the approximate posterior distributions for the respective parameters, and should  be chosen in a way that facilitate the manipulation. In particular, in order to maintain the sparsity enhancing property in the approximate posterior distribution of the
channel BEM coefficients, we choose \cite{VRVM}
\begin{eqnarray}
  Q({\boldsymbol\mu}) &=& \displaystyle\frac{1}{\pi ^M \det(\mathbf{\tilde \Sigma}_{\mu})}
  \exp\{-(\boldsymbol\mu-\mathbf{\tilde m}_{\mu})^H\mathbf{\tilde \Sigma}_{\mu}^{-1}(\boldsymbol\mu-\mathbf{\tilde m}_{\mu})\}\label{qh}\\
  Q(\alpha_j) &=& Ga(\alpha_j|\tilde a_j, \tilde b_j) = \tilde b_j ^{\tilde a_j}\alpha_j^{\tilde a_j -1}\exp(-\tilde b_j\alpha_j)/\Gamma(\tilde a_j)\label{qalpha}
\end{eqnarray}
with $\mathbf{\tilde m}_{\mu}$, $\mathbf{\tilde \Sigma}_{\mu}$ $\tilde a_j$ and $\tilde b_j$ being unknown parameters. Furthermore, for composite noise power, we set $Q(\beta)$ as
\begin{equation}
  Q(\beta) = Ga(\beta|\tilde c,\tilde d) = \tilde d^{\tilde c}\beta^{\tilde c-1}\exp(-\tilde d\beta)/\Gamma(\tilde c)\label{qbeta}
\end{equation}
with $\tilde c$ and $\tilde d$ being unknown parameters. And for $\mathbf{x}_d$, in view of its discrete property, a close approximation is
given as \cite{VI2}
\begin{equation}
  Q(\mathbf{x}_d) = \delta(\mathbf{x}_d-\mathbf{\tilde x}_d)\label{qxd},
\end{equation} with $\mathbf{\tilde x}_d$ being a parameter of $Q(\mathbf{x}_d)$.

With all the distribution functions given above, the nine terms in (\ref{fe}) can be computed respectively. The detailed calculations are shown in the Appendix \ref{A}. With the obtained results (\ref{1}), (\ref{2}), (\ref{3}), (\ref{4}), (\ref{5}), (\ref{6}), (\ref{7}), (\ref{8}), (\ref{9}), and after eliminating some constant terms, the closed-form expression of the free energy function can be written as
\begin{eqnarray}\label{closefe}
  &&\mathbb{F}(\mathbf{\tilde m}_{\mu}, \mathbf{\tilde \Sigma}_{\mu}, \tilde a_j, \tilde b_j, \tilde c, \tilde d, \mathbf{\tilde x}_d)\nonumber \\
  &=& -\log\det(\mathbf{\tilde \Sigma}_{\mu})
        + \mathrm{Tr}\Big\{\mathrm{diag}\big\{\big[
        \frac{\tilde a_1}{\tilde b_1},\ldots, \frac{\tilde a_M}{\tilde b_M}\big]\big\}
        (\mathbf{\tilde m}_{\mu}\mathbf{\tilde m}_{\mu}^H + \mathbf{\tilde \Sigma}_{\mu})\Big\}\nonumber\\
  && + \sum_{j=1}^M\Big[\tilde a_j\log\tilde b_j + (\tilde a_j-1)[\Psi(\tilde a_j) -\log\tilde b_j] -\tilde a_j -\log\Gamma(\tilde a_j)\Big]\nonumber\\
  && -\sum_{j=1}^M\Big[a_j\log b_j + (a_j-1)[\Psi(\tilde a_j) -\log\tilde b_j] -b_j\tilde a_j/\tilde b_j -\log\Gamma(a_j)\Big]\nonumber\\
  && -\sum_{j=1}^M\Big[\Psi(\tilde a_j) -\log\tilde b_j\Big] +\tilde c\log\tilde d + (\tilde c-1)[\Psi(\tilde c) -\log\tilde d] -\tilde c \nonumber\\
  && -\log\Gamma(\tilde c) - (c-1)[\Psi(\tilde c) -\log\tilde d] + d\tilde c/\tilde d -N[\Psi(\tilde c) -\log\tilde d]\nonumber\\
  && + \frac{\tilde c}{\tilde d}\left[ \mathrm{Tr}
  \left\{\mathbf{G}^H[\mathbf{\tilde x}]
  \mathbf{G}[\mathbf{\tilde x}]
  (\mathbf{\tilde m}_{\mu}\mathbf{\tilde m}_{\mu}^H + \mathbf{\tilde \Sigma}_{\mu})\right\}
  +\mathbf{y}^H\mathbf{y} -2\mathrm{Re}\left\{\mathbf{y}^H\mathbf{G}[\mathbf{\tilde x}]
  \mathbf{\tilde m}_{\mu}\right\}\right]\nonumber\\
  &&+\sum_{n=1}^{N_d}\log\Big\{\sum_{{\bar x}_d(n)\in \mathbb{C}_d}\delta({\tilde x}_d(n)-{\bar x}_d(n))\Big\},
\end{eqnarray}
where $\Psi(a) = \frac{\partial}{\partial a}\log \Gamma(a)$ is the digamma function, and we let $\mathbf{\tilde x} = \mathbf{E}_p\mathbf{x}_p + \mathbf{E}_d\mathbf{\tilde x}_d$ to simplify the expression.
Notice that, the free energy function only depends on $\mathbf{\tilde m}_{\mu}$, $\mathbf{\tilde \Sigma}_{\mu}$, $\tilde a_j$, $\tilde b_j$, $\tilde c$, $\tilde d$ and $\mathbf{\tilde x}_d$.
\subsection{Iterative Minimization of Free Energy Function}
After obtaining the closed-form free energy function (\ref{closefe}), the next step is
to minimize the free energy function in order to obtain the optimal $\mathbf{\tilde m}_{\mu},
\mathbf{\tilde \Sigma}_{\mu}, \tilde a_j, \tilde b_j, \tilde c, \tilde d$ and $\mathbf{\tilde x}_d$.
As the function depends on a large number of parameters, it is difficult to  obtain the optimal parameters analytically in one step. The
commonly used solution is to update each one in turn.
In the following, it is shown that the closed-form solutions of  $\mathbf{\tilde m}_{\mu}$ and
$\mathbf{\tilde \Sigma}_{\mu}$ can be derived if the other parameters are fixed. Since the $Q({\boldsymbol\mu})$
is assumed to be in Gaussian form, the optimal BEM coefficient estimate is equal to the mean, i.e.,
$\mathbf{\tilde m}_{\mu}$. Similarly, it is shown below that $(\tilde a_j, \tilde b_j)$ and $(\tilde c, \tilde d)$ can be updated in pairs. Furthermore, we can derive the optimal $\mathbf{\tilde x}_d$ with Viterbi algorithm when other parameters are fixed. Therefore,
$\mathbb{F}(\mathbf{\tilde m}_{\mu}, \mathbf{\tilde \Sigma}_{\mu}, \tilde a_j, \tilde b_j, \tilde c,
 \tilde d, \mathbf{\tilde x}_d) $ is minimized iteratively, starting with a certain
 initial value, and is guaranteed to converge \cite{Bishop}.
\subsubsection{{{Minimization w.r.t. $\mathbf{\tilde x}_d$}}}
Gathering the terms in (\ref{closefe}) that involve $\mathbf{\tilde x}_d$, we have
\begin{eqnarray}\label{ofxd}
\mathbb{F}_{\tilde x_d} &=& \frac{\tilde c}{\tilde d}\left[ \mathrm{Tr}\left\{\mathbf{G}^H[\mathbf{\tilde x}]
\mathbf{G}[\mathbf{\tilde x}](\mathbf{\tilde m}_{\mu}\mathbf{\tilde m}_{\mu}^H +
\mathbf{\tilde \Sigma}_{\mu})\right\}-2\mathrm{Re}\{\mathbf{y}^H\mathbf{G}[\mathbf{\tilde x}]\mathbf{\tilde m}_{\mu}\}\right]\nonumber\\
&&+\sum_{n=1}^{N_d}\log\Big\{\sum_{{\bar x}_d(n)\in \mathbb{C}_d}\delta({\tilde x}_d(n)-{\bar x}_d(n))\Big\}.
\end{eqnarray}
Instead of treating $\mathbf{\tilde x}_d$ as continuous and taking derivatives of (\ref{ofxd}), which does not guarantee the resulted $\mathbf{\tilde x}_d$ will fall on the pre-defined constellation points, optimal $\mathbf{\tilde x}_d$ is searched on
the constellation to minimize (\ref{ofxd}) as follows.

First, it should be noticed that if we only search on the constellation points, then $\tilde x_d(n)\in\mathbb{C}_d$ and $\sum_{{\bar x}_d(n)\in \mathbb{C}_d}\delta({\tilde x}_d(n)-{\bar x}_d(n)) =1$
for all $n$. Thus
\begin{equation}
\sum_{n=1}^{N_d}\log\Big\{\sum_{{\bar x}_d(n)\in \mathbb{C}_d}\delta({\tilde x}_d(n)-{\bar x}_d(n))\Big\}
=\sum_{n=1}^{N_d}\log\{1\} = 0.
\end{equation}
Moreover, ${\tilde c},{\tilde d}>0$ from the property of Gamma distribution, then the factor $\tilde c/\tilde d$ can be excluded in the searching metric, and we  have
\begin{eqnarray}\label{ofxd2}
  \mathbb{F}_{\tilde x_d} &=& \mathrm{Tr}\left\{\mathbf{G}^H[\mathbf{\tilde x}]
\mathbf{G}[\mathbf{\tilde x}](\mathbf{\tilde m}_{\mu}\mathbf{\tilde m}_{\mu}^H +
\mathbf{\tilde \Sigma}_{\mu})\right\} -2\mathrm{Re}\{\mathbf{y}^H\mathbf{G}[\mathbf{\tilde x}]\mathbf{\tilde m}_{\mu}\}.
\end{eqnarray}

\begin{figure}[!htb]
\begin{center}
  \includegraphics[width=.35\textwidth]{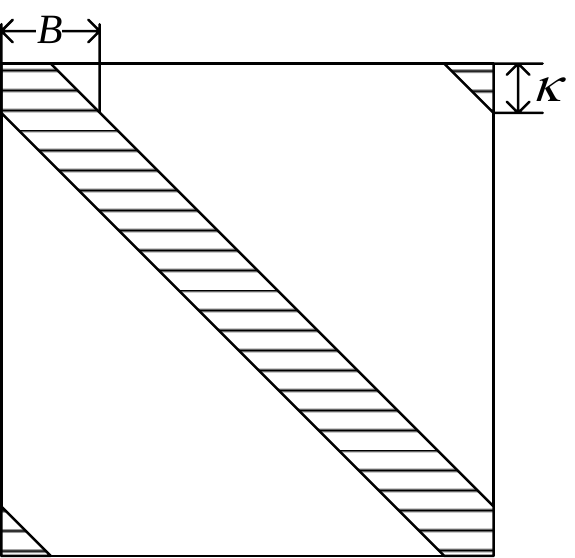}
  \caption{Banded Matrix Structure Approximation for $\mathbf{D}[\mathbf{\tilde m}_{\mu}]$ and $\mathbf{D}[\boldsymbol\xi_j]$} \label{bandlimitedmatrix}
\end{center}
\end{figure}

It should be noticed that, the objective function given in (\ref{ofxd2}) depends on $\mathbf{\tilde x}_d$ in a highly
nonlinear way, making it difficult to find a solution for the optimal $\mathbf{x}_d$.
In order to proceed, we perform the eigen-decomposition $\mathbf{\tilde \Sigma}_{\mu} = \sum_{j= 1}^M \lambda_j \mathbf{\boldsymbol\xi}_j\mathbf{\boldsymbol\xi}_j^H$, and we have
\begin{eqnarray}\label{decompo}
\mathrm{Tr}\left\{\mathbf{G}^H[\mathbf{\tilde x}]\mathbf{G}[\mathbf{\tilde x}]\mathbf{\tilde \Sigma}_{\mu}\right\} =& \sum_{j= 1}^M \lambda_j \mathbf{\boldsymbol\xi}_j^H\mathbf{G}^H[\mathbf{\tilde x}]\mathbf{G}[\mathbf{\tilde x}]\mathbf{\boldsymbol\xi}_j.
\end{eqnarray}
Putting (\ref{decompo}) into (\ref{ofxd2}), we obtain
\begin{equation}\label{ofxd3}
\mathbb{F}_{\tilde x_d} =  \mathbf{\tilde m}_{\mu}^H\mathbf{G}^H[\mathbf{\tilde x}]\mathbf{G}[\mathbf{\tilde x}]\mathbf{\tilde m}_{\mu}
 +\sum_{j= 1}^M \lambda_j \mathbf{\boldsymbol\xi}_j^H\mathbf{G}^H[\mathbf{\tilde x}]\mathbf{G}[\mathbf{\tilde x}]\mathbf{\boldsymbol\xi}_j-2\mathrm{Re}\{\mathbf{y}^H\mathbf{G_h[\tilde x]\tilde m}_{\mu}\}.
\end{equation}
Due to the equality $\mathbf{G[x]\boldsymbol\mu=D[\boldsymbol\mu]x}$ derived from (\ref{SM1}) and (\ref{SM2}), (\ref{ofxd3}) can be written as
\begin{equation}\label{fxd}
    \mathbb{F}_{\tilde x_d} =
    \mathbf{\tilde x}^H\mathbf{D}^H[\mathbf{\tilde m}_{\mu}]\mathbf{D}[\mathbf{\tilde m}_{\mu}]\mathbf{\tilde x}
 +\sum_{j= 1}^M \lambda_j \mathbf{\tilde x}^H\mathbf{D}^H[\boldsymbol{\boldsymbol\xi}_j]\mathbf{D}[\boldsymbol{\boldsymbol\xi}_j]\mathbf{\tilde x}-2\mathrm{Re}\{\mathbf{y}^H\mathbf{D}[\mathbf{\tilde m}_{\mu}]\mathbf{\tilde x}\}.
\end{equation}
Then we adopt the Viterbi algorithm \cite{VAEqualizer} to minimize (\ref{fxd}).
The frequency domain signal $\mathbf{\tilde x}$ is treated as a sequence of data and the correlation between data is determined by the ICI. Strictly speaking, for
DSC, all the elements of $\mathbf{D}[\mathbf{\tilde m}_{\mu}]$ and $\mathbf{D}[\boldsymbol\xi_j]$,
$j = 1,\ldots, M$ are nonzero.
But the entries close to the main diagonal are more prominent compared to those further
away from the diagonal. This means that, the matrices $\mathbf{D}[\mathbf{\tilde m}_{\mu}]$ and $\mathbf{D}[\boldsymbol\xi_j]$ can be approximated as a banded matrix of bandwidth $B=2\kappa+1$, as shown in Figure \ref{bandlimitedmatrix}. This approximated banded structure of DSC matrix considers the most significant ICI from the left and right $\kappa$ closest symbols, and has been widely used in the literature \cite{ModelReduction,VAEqualizer}. As shown in
Figure \ref{bandlimitedmatrix}, the upper right and lower left nonzero entries of the matrix cause cyclic interference between the first and last subcarriers, making the problem
different from the uni-directional convolutional code decoding, where Viterbi algorithm
is commonly used. In order to avoid the complication, we set the first and last $\kappa$ symbols to be zero.

With the banded structure of $\mathbf{D}[\mathbf{\tilde m}_{\mu}]$ and $\mathbf{D}[\boldsymbol\xi_j]$,
the branch metric becomes
\begin{eqnarray}\label{Va}
\mathcal{V}_B([\mathbf{\tilde x}]_{n-2\kappa:n};y(n-\kappa))&= & ([\mathbf{D}[\mathbf{\tilde m}_{\mu}]]_{n,n-\kappa:n-2\kappa}[\mathbf{\tilde x}]_{n-2\kappa:n})^H
([\mathbf{D}[\mathbf{\tilde m}_{\mu}]]_{n,n-\kappa:n-2\kappa}[\mathbf{\tilde x}]_{n-2\kappa:n}) \nonumber\\
&&+\sum_{j= 1}^M \lambda_j ([\mathbf{D}[\boldsymbol{\boldsymbol\xi}_j]]_{n,n-\kappa:n-2\kappa}
[\mathbf{\tilde x}]_{n-2\kappa:n})^H([\mathbf{D}[\boldsymbol{\boldsymbol\xi}_j]]_
{n,n-\kappa:n-2\kappa}
[\mathbf{\tilde x}]_{n-2\kappa:n})\nonumber\\
&&-2\mathrm{Re}\{{y}^*(n-{\kappa})[\mathbf{D}[\mathbf{\tilde m}_{\mu}]]_{n,n-\kappa:n-2\kappa}[\mathbf{\tilde x}]
_{n-2\kappa:n}\}.
\end{eqnarray}
Since known pilot symbols are inserted between unknown data, slight modification to
standard Viterbi algorithm is needed. For $n\in \mathfrak{I}_p$, the true pilot
value will be used to calculate the branch value, and the number of remaining paths will
reduce by half because of merging. And for $n\in \mathfrak{I}_d$, the algorithm will perform as
normal Viterbi algorithm.\subsubsection{{Minimization w.r.t. $\mathbf{\tilde m}_{\mu}, \mathbf{\tilde \Sigma}_{\mu}, \tilde a_j, \tilde b_j, \tilde c$ and $\tilde d$}}
The optimal values of other unknown parameters are obtained by setting the first order
derivative of (\ref{closefe}) with respect to the corresponding parameter to zero.
As shown in Appendix \ref{B}, we obtain the following set of
solutions:
\begin{eqnarray}
\mathbf{\tilde \Sigma}_{\mu} &=& \left(\mathrm{diag}\left\{\left[\frac{\tilde a_1}{\tilde b_1},\ldots, \frac{\tilde a_M}{\tilde b_M}\right]\right\} + \frac{\tilde c}{\tilde d}\mathbf{G}^H[\mathbf{\tilde x}]\mathbf{G}[\mathbf{\tilde x}]\right)^{-1}\label{o1}\\
\mathbf{\tilde m}_{\mu} &=&  \frac{\tilde c}{\tilde d}\mathbf{\tilde \Sigma}_{\mu}\mathbf{G}^H[\mathbf{\tilde x}]\mathbf{y}\label{o2}\\
 \tilde a_j &=& a_j+1\label{o3}\\
\tilde b_j &=& b_j + |[\mathbf{\tilde m}_{\mu}]_{j}|^2 + [\mathbf{\tilde \Sigma}_{\mu}]_{j,j}\label{o4}\\
\tilde c &=& c+N\label{o5}\\
\tilde d &=& d + \mathbf{y}^H\mathbf{y} - 2\mathrm{Re}\left\{\mathbf{y}^H\mathbf{G}\mathbf{[\tilde x]\tilde m}_{\mu}\right\}+\mathrm{Tr}\left\{\mathbf{G}^H[\mathbf{\tilde x}]\mathbf{G}[\mathbf{\tilde x}]\left(\mathbf{\tilde m}_{\mu}\mathbf{\tilde m}_{\mu}^H + \mathbf{\tilde \Sigma}_{\mu}\right)\right\}.\label{o6}
\end{eqnarray}
It is worth noting that, along with each update, when $|[\mathbf{\tilde m}_{\mu}]_{j}|^2 + [\mathbf{\tilde \Sigma}_{\mu}]_{j,j}$
gets close to zero, meaning both mean and variance of the corresponding ${\mu_j}$ are close to zero, then ${\mu_j}$ can be treated as null entry and pruned from further iteration. In
practice, a threshold with the order of $10^{-10}$ is used to compare with $|[\mathbf{\tilde m}_{\mu}]_{j}|^2 + [\mathbf{\tilde \Sigma}_{\mu}]_{j,j}$
to determine which ${\mu_j}$ is being pruned~\cite{SBLandRVM}.
\subsection{Summary of the Iterative Algorithm}
We summarize the parameter updating procedure as follows\\
\textbf{Initialization:} Choose initial values $\{\tilde a_1^0,\ldots, \tilde a_M^0\}$,
$\{\tilde b_1^0,\ldots,\tilde b_M^0\}$, $\tilde c^0,\tilde d^0$ and $\mathbf{\tilde x}_d^0$.\\
\textbf{Iterations:} For the $i^{th}$ iteration\\
{\it \underline{Updating the parameters of GCE-BEM coefficients}} 
\begin{eqnarray*}
\mathbf{\tilde \Sigma}_{\mu}^i &=& \Big(\mathrm{diag}\big\{\big[\frac{\tilde a_1^{i-1}}{\tilde b_1^{i-1}},\ldots, \frac{\tilde a_M^{i-1}}{\tilde b_M^{i-1}}\big]\big\} + \frac{\tilde c^{i-1}}{\tilde d^{i-1}}\mathbf{G}^H[\mathbf{\tilde x}^{i-1}]\mathbf{G}[\mathbf{\tilde x}^{i-1}]\Big)^{-1}\\
\mathbf{\tilde m}_{\mu}^i &=&\frac{\tilde c^{i-1}}{\tilde d^{i-1}} \mathbf{\tilde \Sigma}_{\mu}^i
\mathbf{G}^H[\mathbf{\tilde x}^{i-1}]\mathbf{y}
\end{eqnarray*}
{\it \underline{Updating the hyperparameters of GCE-BEM coefficients }} 
\begin{eqnarray*}
  \tilde a_j^i &=& a_j+1\\
\tilde b_j^i &=& b_j + |[\mathbf{\tilde m}_{\mu}]_{j}|^2 + [\mathbf{\tilde \Sigma}_{\mu}]_{j,j}
\end{eqnarray*}
{\it \underline{Updating the estimate of data}} 
 \[\begin{split}\min_{\mathbf{\tilde x}_d}\mathbb{F}_{\tilde x_d}
    = & -2\mathrm{Re}\left\{\mathbf{y}^H\mathbf{D}[\mathbf{\tilde m}_{\mu}]\mathbf{\tilde x}\right\}+
    \mathbf{\tilde x}^H\mathbf{D}^H[\mathbf{\tilde m}_{\mu}]\mathbf{D}[\mathbf{\tilde m}_{\mu}]\mathbf{\tilde x} +\sum_{j= 1}^M \lambda_j \mathbf{\tilde x}^H\mathbf{D}^H[\boldsymbol{\boldsymbol\xi}_j]\mathbf{D}[\boldsymbol{\boldsymbol\xi}_j]\mathbf{\tilde x},
  \end{split}\]
where   $\mathbf{\tilde \Sigma}_{\mu}^i = \sum_{j=1}^{M^{i-1}}\lambda_j^i\mathbf{\boldsymbol\xi}_j^i(\mathbf{\boldsymbol\xi}_j^i)^H$, using Viterbi algorithm.\\
{\it \underline{Updating the hyperparameters of noise}} 
\begin{eqnarray*}
  \tilde c^i &=& c+N\\
\tilde d^i &=& d + \mathbf{y}^H\mathbf{y} - 2\mathrm{Re}\left\{\mathbf{y}^H\mathbf{G}\mathbf{[\tilde x}^i]\mathbf{\tilde m}_{\mu}^i\right\}+\mathrm{Tr}\left\{\mathbf{G}^H[\mathbf{\tilde x}^i]\mathbf{G}[\mathbf{\tilde x}^i]\left(\mathbf{\tilde m}_{\mu}^i(\mathbf{\tilde m}_{\mu}^i)^H + \mathbf{\tilde \Sigma}_{\mu}^i\right)\right\}
\end{eqnarray*}
{\it \underline{Pruning}} \\
If $|[\mathbf{\tilde m}_{\mu}]_{j}|^2 + [\mathbf{\tilde \Sigma}_{\mu}]_{j,j}<10^{-10} \Rightarrow {\mu_j} = 0$, remove
$\tilde a_j$ from $\{\tilde a_1^i,\ldots, \tilde a_M^i\}$ and $\tilde b_j$ from
$\{\tilde b_1^i,\ldots, \tilde b_M^i\}$ and update the expression of $\mathbf{G[\tilde x}^i]$
by removing the $j^{th}$ column.\\
\textbf{End}\\
It is worth noting that the values of $\{\tilde a_1,\ldots, \tilde a_M\}$ and $\tilde c$
remain the same for each iteration, thus they should only be updated in the first iteration
in practice.
\section{Initialization of the Iterative Algorithm}\label{initial}
In order to start the iteration, one of the quantities we need is the initial data estimate $\mathbf{\tilde x}_d^0$. If an initial channel estimate can be obtained from pilots, then from (\ref{SM2}), the initial data estimate can be obtained.
In Section \ref{GCEreform}, GCE-BEM with large oversampling factor is used to represent
the channel. Though this provides flexibility for selecting important bases in the
iterative channel estimation and data detection, the large number of unknown coefficients
corresponding to GCE-BEM also brings challenges in the initial value estimation, which is
important to any iterative algorithm. Unfortunately, traditional compressed sensing methods
like BP and orthogonal matching pursuit (OMP), whose performance highly depends on the
accurate knowledge of noise variance, is not applicable in our case, since the
variance of the composite noise is unknown.

On the other hand, notice that the proposed iterative algorithm does not rely directly on the estimate of ${\boldsymbol\mu}$ to start the iteration. The channel is needed only indirectly through initial data estimate.
Thus, during initial data detection, we choose to expand the channel with CE-BEM, which corresponds to choosing the oversampling factor
of GCE-BEM as $V = 1$.  Then least squares (LS) algorithm is used to obtain
the initial channel estimation. Though CE-BEM is widely reported as relatively inaccurate among other BEMs, it represents the time-varying channel using a very small number of orthogonal bases.
This is highly beneficial for a rough estimate at the initial stage where there is little knowledge of the channel and source signal at the receiver.

According to (\ref{datastructure}), $\mathbf{x = E}_d\mathbf{x}_d + \mathbf{E}_p\mathbf{x}_p$. Together with the fact that $\mathbf{G[E}_p\mathbf{x}_p+\mathbf{E}_d\mathbf{x}_d] = \mathbf{G[E}_p\mathbf{x}_p]+\mathbf{G[E}_d\mathbf{x}_d]$, (\ref{SM1}) can be written as $
    \mathbf{y}=\mathbf{G[E}_p\mathbf{x}_p]{\boldsymbol\mu}+\mathbf{G[E}_d\mathbf{x}_d]{\boldsymbol\mu}+\mathbf{v}$.
Collecting the output samples corresponding to pilot positions $\mathfrak{I}_p$, the equation for initial channel estimation can be written as
\begin{equation}\label{SMpilots}
    \mathbf{y}_p = \mathbf{G}_p\mathbf{[E}_p\mathbf{x}_p]{\boldsymbol\mu}+\mathbf{G}_p\mathbf{[E}_d\mathbf{x}_d]{\boldsymbol\mu}\mathbf{+v}_p,
\end{equation}
where $\mathbf{G}_p[.]$ is constructed from the rows of $\mathbf{G}[.]$ corresponding to $\mathfrak{I}_p$.
The initial channel estimation can be obtained by treating the second term of (\ref{SMpilots}) as noise and
performing LS algorithm:
\begin{equation}\label{h0}
  \hat{{\boldsymbol\mu}} = (\mathbf{G}_p^H\mathbf{[E}_p\mathbf{x}_p] \mathbf{G}_p\mathbf{[E}_p\mathbf{x}_p])^{-1} \mathbf{G}^H_p\mathbf{[E}_p\mathbf{x}_p]\mathbf{y}_p.
\end{equation}

With the estimated channel $\hat{{\boldsymbol\mu}}$, we rewrite (\ref{SM2}) as
$ \mathbf{y= D[\hat{{\boldsymbol\mu}} ]E}_d\mathbf{x}_d + \mathbf{D[\hat{{\boldsymbol\mu}} ]E}_p\mathbf{x}_p + \mathbf{v}.$
Applying LS estimation again, we have
\begin{equation}
  \mathbf{\hat x}_d^0 = (\mathbf{E}_d^H\mathbf{D}^H[\hat{{\boldsymbol\mu}} ]\mathbf{D}[\hat{{\boldsymbol\mu}} ]\mathbf{E}_d )^{-1}\mathbf{E}_d^H\mathbf{D}^H[\hat{{\boldsymbol\mu}} ](\mathbf{y} - \mathbf{D[\hat{{\boldsymbol\mu}} ]E}_p\mathbf{x}_p).
\end{equation}
The obtained $\mathbf{\hat x}_d^0$ may not reside on the constellation map, thus quantization is performed on $\mathbf{\hat x}_d^0$ and the initial data detection is given as $  \mathbf{\tilde x}_d ^0 = \mathrm{Qant}[\mathbf{\hat x}_d^0]$.
Notice that in DSC, the ICI is not negligible, and $\mathbf{G}_p\mathbf{[E}_d\mathbf{x}_d]
{\boldsymbol\mu}\neq \mathbf{0}$, which decreases the accuracy of estimation in (\ref{h0}), and in turns affects the accuracy of initial data detection. This is the reason why an iterative algorithm is necessary.

For other initial values $\{\tilde a_1^0, \ldots, \tilde a_M^0\}$, $\{\tilde b_1^0,\ldots,
\tilde b_M^0\}$, $\tilde c^0, \tilde d^0$ in the iterative algorithm, it is should be noticed that only the ratios $\tilde a_j/\tilde b_j$ and $\tilde c/\tilde d$ are required, thus we only need to specify the initial values of the ratios to start the iteration.
From (\ref{qalpha}) and the property of Gamma distribution, $\tilde a_j/\tilde b_j$ represents
the mean value of $\alpha_j$, which is the inverse variance of
channel GCE-BEM coefficients. Since we have no information about their relative
values, we can set
them to be equal. That is, let $\tilde a_j^0/\tilde b_j^0 = 1/M$ for all $j$.
Furthermore, from (\ref{qbeta}) and the property of Gamma distribution, $\tilde c/\tilde d= \mathbb{E}\{\beta\} = \mathbb{E}\{1/\varpi_v^2\}$.  Therefore the initial
value can be set as ${\tilde c^0}/{\tilde d^0} = {1}/{\hat\varpi_v^2}$, where $\hat\varpi_v^2$ is an estimate of noise power 
\begin{equation}
  \hat\varpi_v^2 = \mathbf{y-G[E}_p\mathbf{x}_p+\mathbf{E}_d\mathbf{\tilde x}_d^0]\hat{{\boldsymbol\mu}} .
\end{equation}
\section{Simulation Results and Discussions}\label{simulation}
In this section, simulation results of dual-hop and three-hop cooperative OFDM systems are provided. In both systems, each OFDM symbol has 128 subcarriers and the length of CP is 8. Carrier frequency is $f_c = 2$GHz and the sample interval is $T_s = 2\mu s$. The
channel $\bar h_{k,\rho}(n,l)$ is generated according to zero-mean complex Gaussian distribution with autocorrelation of the
$l^{th}$ tap given by $\mathbb{E}\{\bar h_{k,\rho}(m,l)\bar h
_{k,\rho}(n,l)\}=\sigma_{k,\rho}^2(l)J_0(2\pi f_{k,\rho}(l)(m-n)T_s)$ \cite{Jakes}, where
$J_0(\cdot)$ represents the zero-order Bessel function of the first kind, and $\sigma_{k,\rho}^2(l)$
is the power of the $l^{th}$ tap. Fourteen pilot clusters are used. The clusters are equal-spaced and interleaved with data subcarriers. In each cluster, one nonzero pilot is guarded by one zero pilot on each side. The nonzero pilots are generated as zero-mean  complex Gaussian random variables with power three times that of data symbols.  And the data is modulated with QPSK of unit power.  The normalized channel mean-square error (MSE) and data detection bit error rate (BER) are plotted to demonstrate the performance. The MSE of the channel estimate at the $i^{th}$ iteration is defined as $
  \mathrm{MSE}^{i} ={\|\mathbf{\hat H}^i - \mathbf{H}\|^2}/{\|\mathbf{H}\|^2}$,
where $\mathbf{\hat H}^i $ is the channel matrix recovered from the GCE-BEM estimate at the $i^{th}$ iteration. The noise power at the relays and destination are set to be the same $\varpi_d^2 = \varpi_{k,\rho}^2,$ for all $k$ and $\rho$.
The signal-to-noise ratio (SNR) in the following figures is defined as
$\mathrm{SNR} = \sigma_s^2/\varpi_d^2$ \cite{AF1}.  The oversampling factor is chosen as $V = 20$ for GCE-BEM.
Each point is obtained by averaging the results over 1,000 runs.

For the dual-hop system, two relaying paths ($K$ = 2) are considered. For both relaying paths ($k=1,2$), the maximal normalized Doppler shifts\footnote[1]{Normalized Doppler shift is defined as
$Nf_dT_s$ with $f_d$ being the Doppler frequency.} are set as 0.05 for the first hop
and 0.15 for the second hop. In the simulation, for each specific channel, one randomly chosen tap has Doppler shift equals the maximum Doppler shift specified above. And for other taps, their Doppler shifts are uniformly drawn within the range from 0 to the maximum Doppler shift.  Both source-relay channels have 2 taps. One of the source-relay channels has tap positions uniformly drawn from $\{0,1, 2\}$ while the other has tap positions uniformly drawn from $\{0,1,2,3, 4\}$. And both relay-destination channels have 2 taps, with the tap positions uniformly drawn from $\{0,1,2, 3\}$ for one channel and from $\{0,1,2 \}$ for the other. All the channels follow exponential power delay profiles normalized to unit power. For the Viterbi equalizer, $\kappa=3$ is chosen.

\begin{figure}[!htb]
  \centering
  \includegraphics[width=.7\textwidth]{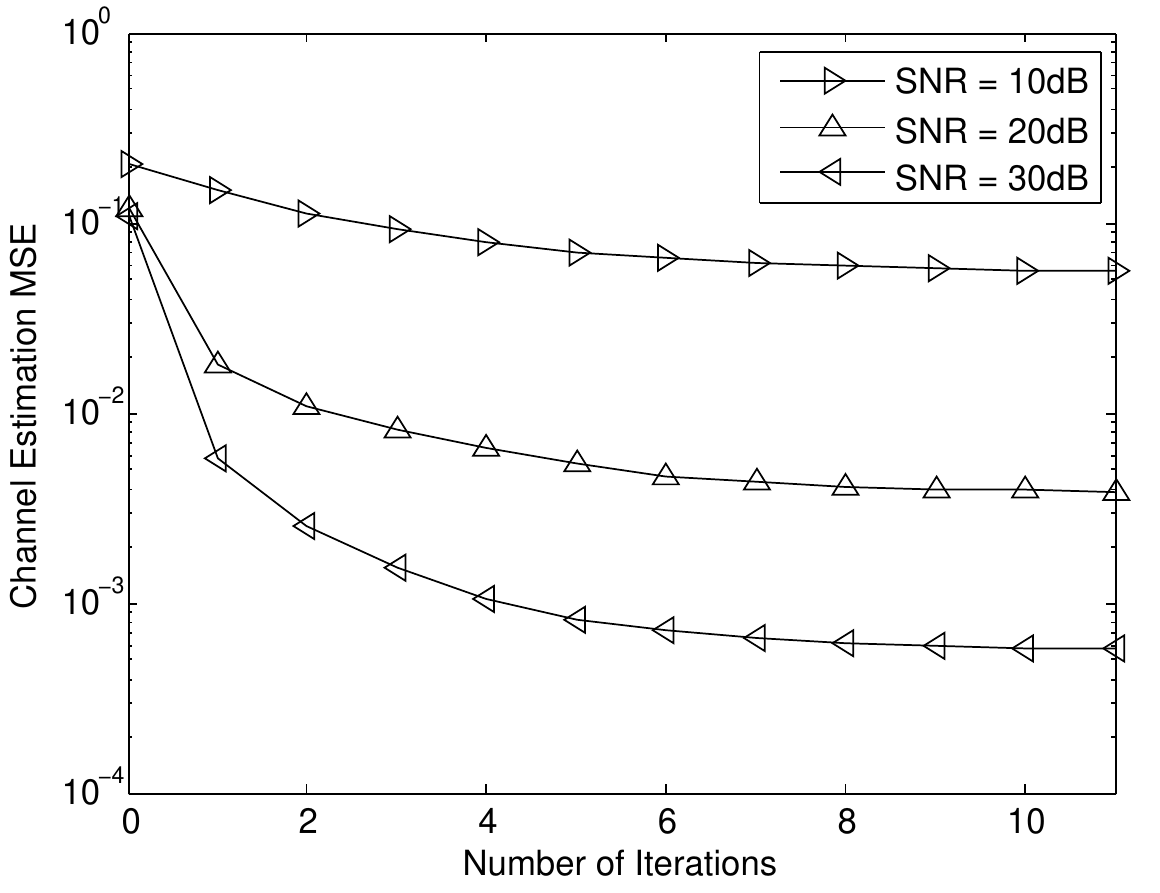}\\
  \caption{Convergence of Channel Estimation for Dualhop OFDM System}\label{RelayMSEConvergence}
\end{figure}

\begin{figure}[!htb]
  \centering
  \includegraphics[width=.7\textwidth]{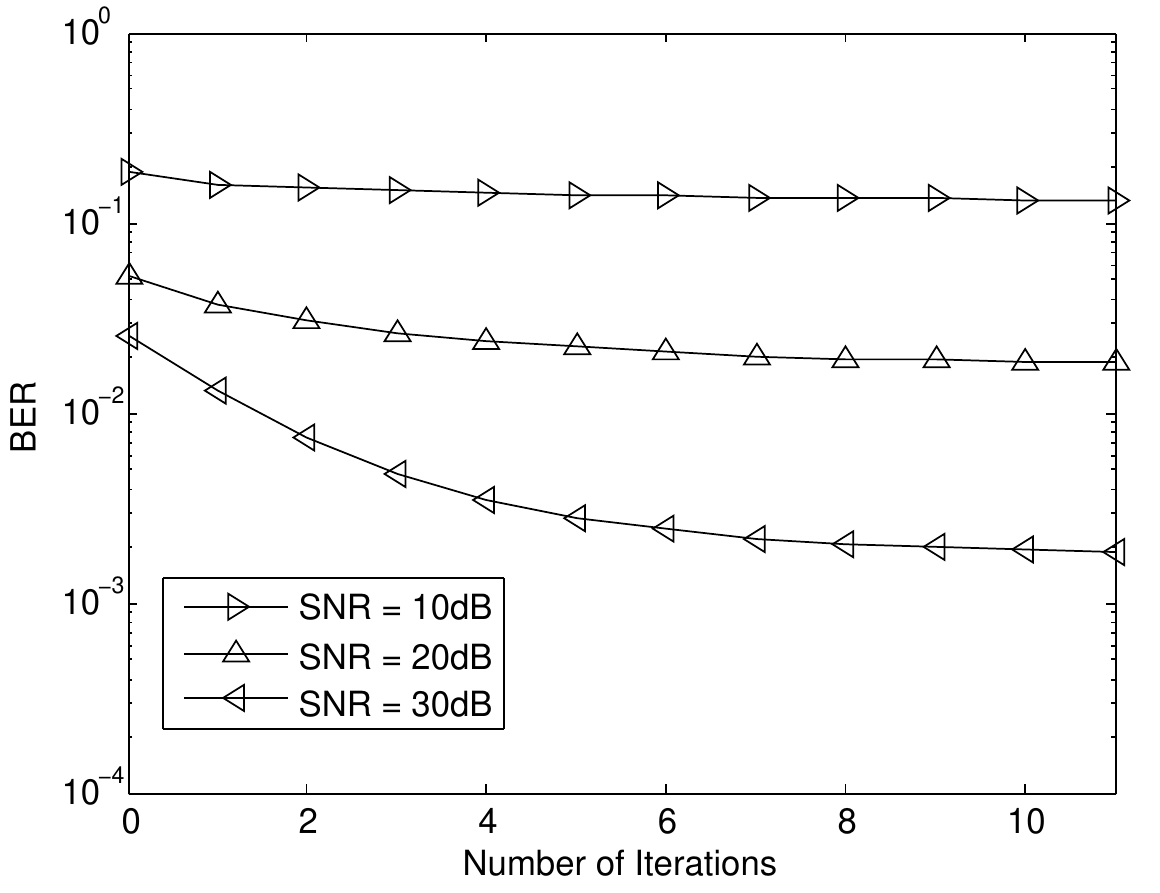}\\
  \caption{Convergence of Data Detection for Dualhop OFDM System}\label{RelayBERConvergence}
\end{figure}

Figure \ref{RelayMSEConvergence} and Figure \ref{RelayBERConvergence} present the convergence performance of the proposed iterative algorithm in terms of MSE and BER, respectively. SNR is set at 10dB, 20dB and 30dB. It can be seen that the MSEs and BERs improve significantly in the first few iterations and converge to stable values before 10 iterations.

\begin{figure}[!htb]
  \centering
  \includegraphics[width=.7\textwidth]{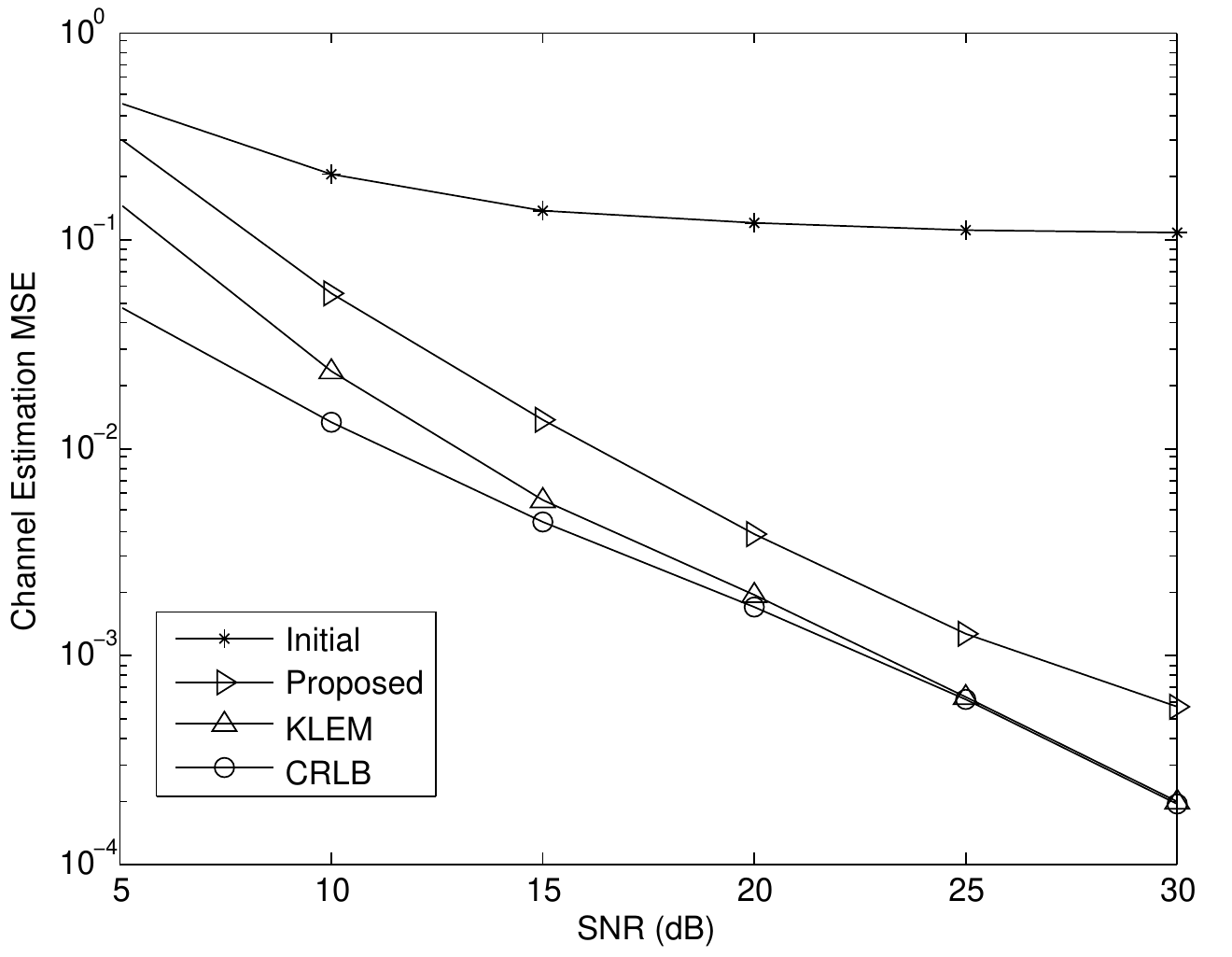}\\
  \caption{Performance of Channel Estimation for Dualhop OFDM System}\label{RelayMSE}
\end{figure}

\begin{figure}[!htb]
  \centering
  \includegraphics[width=.7\textwidth]{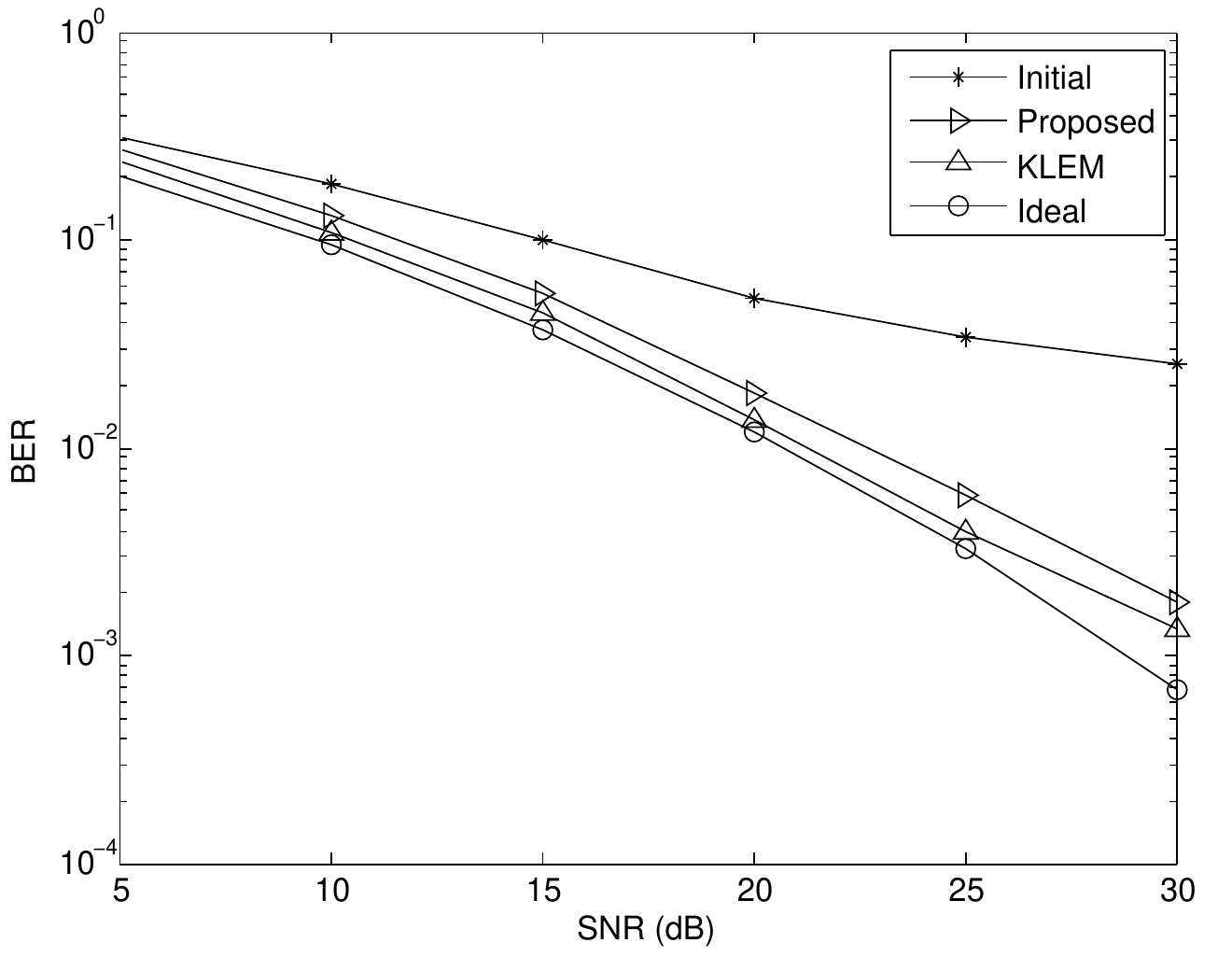}\\
  \caption{Performance of Data Detection for Dualhop OFDM System}\label{RelayBER}
\end{figure}

Figure \ref{RelayMSE} and Figure \ref{RelayBER} show the MSE and BER performance achieved by the proposed iterative algorithm versus SNRs. The results are taken after 10 iterations in order to guarantee convergence. In the figures, KLEM represents the performance of the EM algorithm with channel expanded on Karhuen-Lo\`{e}ve (KL) bases. This algorithm is an extension of the KLEM algorithm for single-hop case \cite{KLEM}, and the detail is not included in this paper due to space limitation. The KLEM algorithm requires full information on channel tap positions, Doppler frequencies and power profile of each channel, together with noise statistics, thus serves as a reference for optimal performance here. And CRLB curve represents the Cram$\mathrm{\acute{e}}$-Rao lower bound, which can be obtained from \cite{KLEM} by replacing the single-hop channel and noise power with the composite channel and composite noise power. Meanwhile, ideal case with full channel information at the receiver is also depicted as the performance bound in the BER figure. From Figure \ref{RelayMSE} and \ref{RelayBER}, it can be seen that, the proposed iterative algorithm successfully eliminates the interdependence between data detection and channel estimation, and exhibits significant performance improvement compared to the initial channel estimation MSE and data detection BER. Furthermore, though the proposed iterative algorithm does not have access to the relaying system structure (number of available links $K$) or any statistical information of the channel and noise, there are only minor performance gaps between the proposed method and the KLEM. Furthermore, the proposed algorithm and KLEM are very close to the ideal data detection in terms of BER performance.

For three-hop relaying system, the maximal normalized Doppler shifts for the first and third hop are set as 0.05 while that of the second hop is set as 0.15. Two relaying paths are considered ($K$= 2). For the first relaying path, the number of channel taps in the three
hops are $\{2,3,2\}$, respectively; while that for the second relaying path is $\{3,2,2\}$,
respectively. The channel taps in each hop are consecutive. For the Viterbi equalizer, $\kappa=4$ is chosen.

\begin{figure}[!htb]
  \centering
  \includegraphics[width=.7\textwidth]{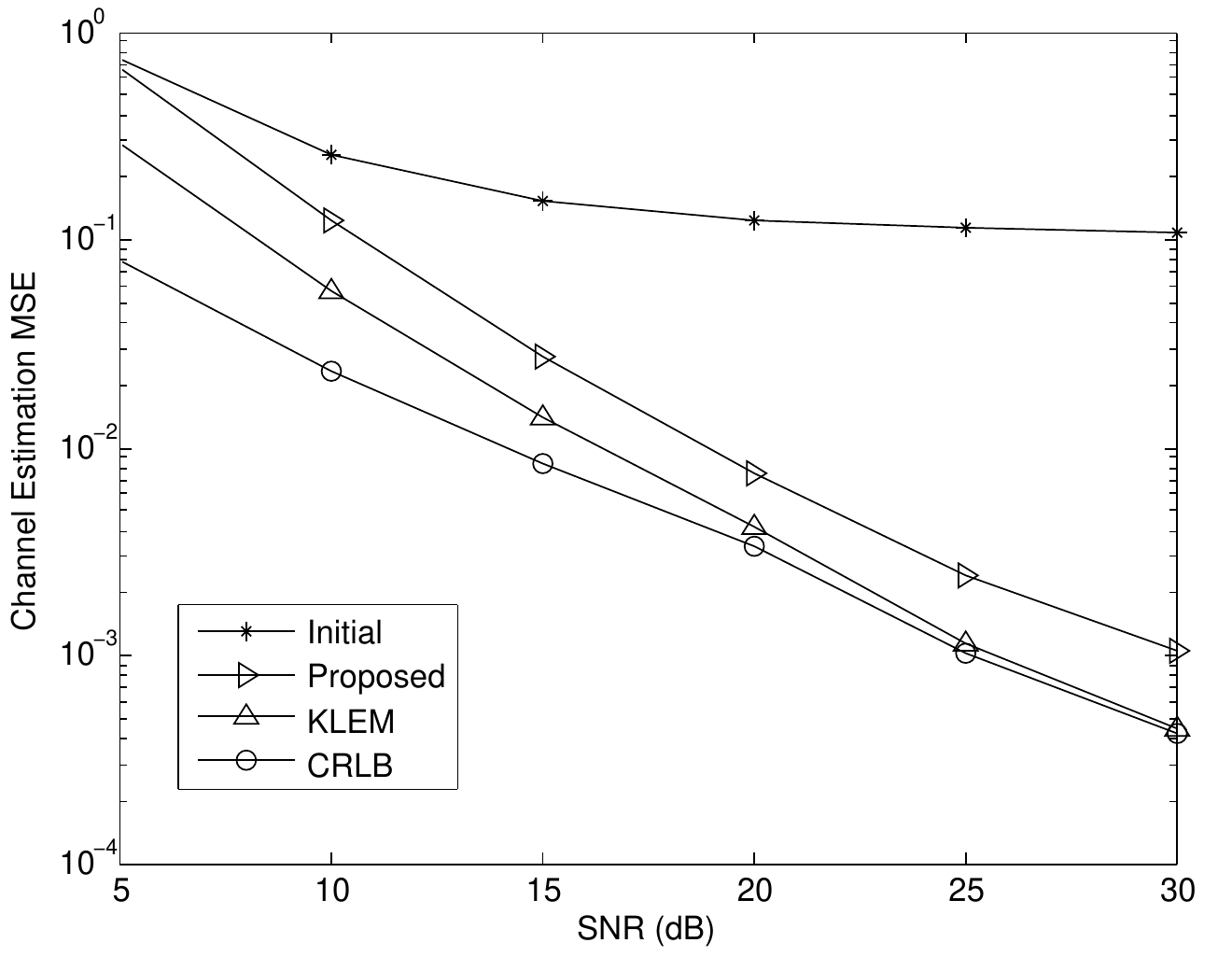}\\
  \caption{Performance of Channel Estimation for Three-hop OFDM System}\label{Helen3hopMSE1000}
\end{figure}

\begin{figure}[!htb]
  \centering
  \includegraphics[width=.7\textwidth]{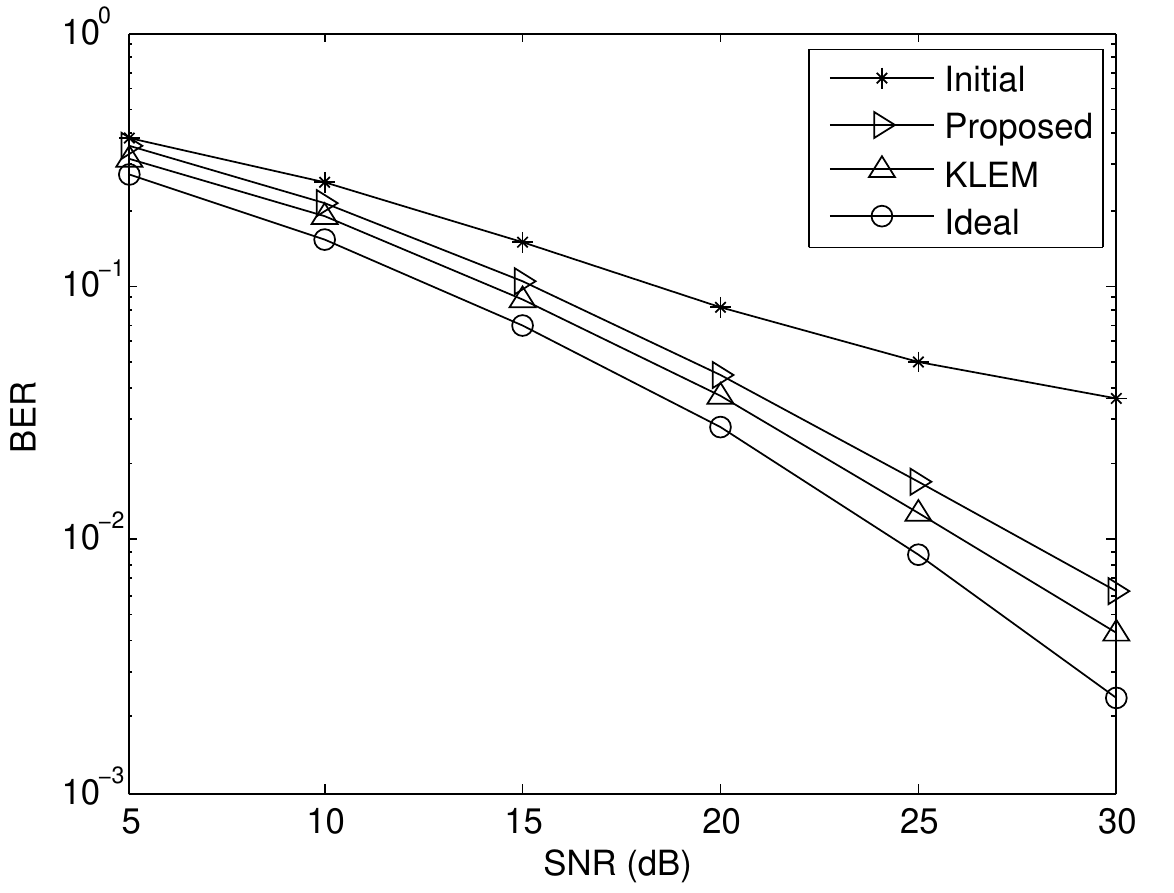}\\
  \caption{Performance of Data Detection for Three-hop OFDM System}\label{Helen3hopBER1000}
\end{figure}

Figure \ref{Helen3hopMSE1000} and Figure \ref{Helen3hopBER1000} show the MSE and BER performance achieved by the proposed iterative algorithm versus SNRs, with results taken after 10 iterations\footnote[2]{As the convergence performance of a three-hop system is similar to that of dual-hop system, the convergence figures are not shown here.}. In both figures, performance curves of KLEM, which demands detailed information of relaying system structure, channel and noise statistics, are depicted as a reference for optimal channel estimation and data detection. From Figure \ref{Helen3hopMSE1000}, it is seen that, the proposed iterative algorithm greatly improve the performance from the initial channel estimation, indicating the ability of the proposed algorithm to cancel interference between unknown data and pilots through iterations. Furthermore, after convergence,
only a small performance gap exists between the proposed algorithm and KLEM, which touches the CRLB at high SNRs. This exhibits the strong ability of our proposed algorithm in learning the statistics of both channel and noise. From Figure \ref{Helen3hopBER1000}, the BER performance of our proposed method is also shown to improve significantly compared to the initial data detection and is very close that of KLEM algorithm. From these figures, it can be concluded that, though the system model in a three-hop system is more complicated than that of dual-hop case, the proposed algorithm continues to present good performance in terms of both channel estimation MSE and data detection BER and demonstrate robustness in a variety of OFDM relaying systems.

\section{Conclusions}\label{conclusion}
In this paper, channel estimation and data detection for multihop OFDM relaying system under high mobility has been investigated with focus on unknown channel orders and Doppler frequencies.
By exploring the matrix structure of channels in different hops, we first simplified the multihop multilink channel matrix into a composite channel matrix. Then the composite channel was represented using GCE-BEM with a large oversampling factor so that sparsity on the delay-Doppler domain was introduced. Sparsity enhancing Gaussian priors with Gamma hyperpriors were adopted to enable the identification of nonzero entries. A pilot-aided iterative algorithm was developed under variational inference (VI) framework, using only limited number of pilot subcarriers in one OFDM symbol. The proposed algorithm iteratively estimates the channel, recovers the unknown data using Viterbi algorithm and learns the channel and noise statistical information.
Simulation results showed that, even without any specific information on system structure, channel tap positions, channel lengths, Doppler shifts and noise powers, the proposed algorithms exhibited performance very close to that of an optimal channel estimation and data detection algorithm, which requires all of the above information.

\appendices
\section{Calculation of Free Energy Function}\label{A}

Taking logarithm on (\ref{qh}) and substituting the result to the  \emph{\textbf{{first term}}} of (\ref{fe}), we obtain
\begin{eqnarray}\label{1}
  \int_{{\boldsymbol\mu}} Q({\boldsymbol\mu}) \log Q({\boldsymbol\mu})d{\boldsymbol\mu} &=& -M\log\pi - \log\det(\mathbf{\tilde \Sigma}_{\mu})-\mathbf{\tilde m}_{\mu}^H \mathbf{\tilde \Sigma}_{\mu}^{-1}\mathbf{\tilde m}_{\mu} \nonumber\\
  &&  +2\mathrm{\mathrm{Re}}\left\{\mathbb{E}\{{\boldsymbol\mu}^H\}\mathbf{\tilde \Sigma}_{\mu}^{-1}\mathbf{\tilde m}_{\mu}\right\} - \mathrm{Tr}\left\{\mathbf{\tilde \Sigma}_{\mu}^{-1} \mathbb{E}\{ \boldsymbol\mu\boldsymbol\mu^H\}\right\}\nonumber\\
  &=& -M\log\pi - \log\det(\mathbf{\tilde \Sigma}_{\mu})-\mathbf{\tilde m}_{\mu}^H \mathbf{\tilde \Sigma}_{\mu}^{-1}\mathbf{\tilde m}_{\mu} \nonumber\\
  &&+ 2\mathbf{\tilde m}_{\mu}^H\mathbf{\tilde \Sigma}_{\mu}^{-1}\mathbf{\tilde m}_{\mu}
  -\mathrm{Tr}\left\{\mathbf{\tilde \Sigma}_{\mu}^{-1} (\mathbf{\tilde m}_{\mu}\mathbf{\tilde m}_{\mu}^H + \mathbf{\tilde \Sigma}_{\mu})\right\}\nonumber\\
  &=& -M\log\pi - \log\det(\mathbf{\tilde \Sigma}_{\mu}) - M.
\end{eqnarray}

On the other hand, from the form of $Q(\alpha_j)$ given in (\ref{qalpha}), and notice that $Q(\boldsymbol\alpha)=\prod _{j=1} ^M Q(\alpha_j)$, we can compute the \emph{\textbf{second term}} of (\ref{fe}) as
\begin{eqnarray}\label{2}
    \int_{\boldsymbol\alpha}Q(\boldsymbol\alpha)\log Q(\boldsymbol\alpha)d\boldsymbol\alpha &= & \sum_{j=1}^M\left[\tilde a_j\log \tilde b_j - \log\Gamma(\tilde a_j) + (\tilde a_j-1)\mathbb{E}\{\log\alpha_j\} -\tilde b_j\mathbb{E}\{\alpha_j\}\right]\nonumber\\
    &=& \sum_{j=1}^M \Big[\tilde a_j\log \tilde b_j - \log\Gamma(\tilde a_j)  \nonumber\\
    &&  + (\tilde a_j-1)\left(\Psi(\tilde a_j)-\log\tilde b_j\right)-\tilde b_j \left({\tilde a_j}/{\tilde b_j}\right) \Big]\nonumber\\
    &=& \sum_{j=1}^M\left[\log\tilde b_j - \log\Gamma(\tilde a_j) + (\tilde a_j-1)\Psi(\tilde a_j) -\tilde a_j\right],
\end{eqnarray}
where the digamma function $\Psi$ is defined by  $\displaystyle\Psi(a) = \frac{\partial}{\partial a}\log\Gamma(a).$ Furthermore, since $Q(\beta)$ is in the same form as $Q(\alpha_j)$, following a similar derivation as above, it can be easily shown that the \emph{\textbf{third term}} of (\ref{fe}) is
\begin{eqnarray}\label{3}
  \int_{\beta} Q(\beta)\log Q(\beta)d\beta = \log\tilde d - \log\Gamma(\tilde c) + (\tilde c-1)\Psi(\tilde c) - \tilde c.
\end{eqnarray}

Based on the Dirac delta function in (\ref{qxd}), the \emph{\textbf{forth term}} of (\ref{fe}) is given by
\begin{eqnarray}\label{4}
 \int_{\mathbf{x}_d}Q(\mathbf{x}_d)\log Q(\mathbf{x}_d)d\mathbf{x}_d
 &=& \log Q(\mathbf{\tilde x}_d)= \log \delta(\mathbf{\tilde x}_d-\mathbf{\tilde x}_d)=0.
\end{eqnarray}
Furthermore, from (\ref{hprior}), we have
\begin{eqnarray}
\log p({\boldsymbol\mu}|\boldsymbol \alpha) &=& -M\log\pi -\log\det(\mathbf{A}^{-1}) - {\boldsymbol\mu}^H\mathbf{A}{\boldsymbol\mu}\nonumber\\
&=& -M\log\pi +\log\left(\prod_{j=1}^M \alpha_j\right) - \mathrm{Tr}\left\{\mathbf{A}{\boldsymbol\mu}{\boldsymbol\mu}^H\right\}.
\end{eqnarray}
And the \emph{\textbf{fifth term}} of (\ref{fe}) can be computed as
\begin{eqnarray}\label{5}
    &&\int_{{\boldsymbol\mu},\boldsymbol\alpha} Q({\boldsymbol\mu})Q(\boldsymbol\alpha)\log p({\boldsymbol\mu}|\boldsymbol\alpha)d{\boldsymbol\mu}d\boldsymbol\alpha \nonumber\\
    &=&
    -M\log\pi +\sum_{j=1}^M\mathbb{E}_{\alpha}\{\log\alpha_j\}-\mathrm{Tr}\{\mathbb{E}_{\alpha}
    \{\mathrm{diag}\{\boldsymbol\alpha\}\}\mathbb{E}_{\mu}\{{\boldsymbol\mu}{\boldsymbol\mu}^H\}\}\nonumber\\
    &=& -M\log\pi + \sum_{j=1}^M\left(\Psi(\tilde a_j) -\log\tilde b_j\right)-\mathrm{Tr}\bigg\{\mathrm{diag}\left\{\left[\frac{\tilde a_1}{\tilde b_1}, \ldots, \frac{\tilde a_M}{\tilde b_M}\right]\right\}\left(\mathbf{\tilde m}_{\mu}\mathbf{\tilde m}_{\mu}^H + \mathbf{\tilde \Sigma}_{\mu}\right)\bigg\}.
\end{eqnarray}

From (\ref{prioralpha}), we can compute the logarithm of $p(\boldsymbol\alpha)=\prod _{j=1} ^M p(\alpha_j)$, and the \emph{\textbf{sixth term}} of (\ref{fe}) can be written as
\begin{eqnarray}\label{6}
    \int_{\boldsymbol\alpha} Q(\boldsymbol\alpha)\log p(\boldsymbol\alpha)d\boldsymbol\alpha
    &=& \sum_{j=1}^M \big[a_j\log b_j - \log\Gamma(a_j)+(a_j-1)\mathbb{E}_{\alpha}\{\log\alpha_j\} -b_j\mathbb{E}\{\alpha_j\}\big]\nonumber\\
    &=& \sum_{j=1}^M \big[a_j\log b_j - \log\Gamma(a_j)+(a_j-1)(
    \Psi(\tilde a_j) -\log\tilde b_j) -b_j\frac{\tilde a_j}{\tilde b_j}\big].
\end{eqnarray}
Similarly, as $Q(\beta)$ and $p(\beta)$ are in the same form as $Q(\alpha_j)$ and $p(\alpha_j)$, we can easily show that the \emph{\textbf{seventh term}} of  (\ref{fe}) is
\begin{equation}\label{7}
  \int_{\beta}Q(\beta)\log p(\beta)d\beta = c\log d -\log\Gamma(c) +(c-1)\left(\Psi(\tilde c)-\log\tilde d\right)-d\frac{\tilde c}{\tilde d}.
\end{equation}

From (\ref{pxd}), we can obtain the logarithm of $p(\mathbf{x}_d)$. Together with (\ref{qxd}),
the \emph{\textbf{eighth term}} of (\ref{fe}) can be derived as
\begin{eqnarray}\label{8}
    \int_{\mathbf{x}_d}Q(\mathbf{x}_d)\log p(\mathbf{x}_d)d\mathbf{x}_d
    &=& \int_{\mathbf{x}_d}\delta(\mathbf{x}_d-\mathbf{\tilde x}_d)\sum_{n=1}^{N_d}\log\big\{\sum_{{\bar x}_d(n)\in \mathbb{C}_d}\delta({x}_d(n)-{\bar x}_d(n))\big\}d \mathbf{x}_d-\log\{\mathcal{M}_d^{N_d}\}\nonumber\\
    &=& \sum_{n=1}^{N_d}\log\Big\{\sum_{{\bar x}_d(n)\in \mathbb{C}_d}\delta({\tilde x}_d(n)-{\bar x}_d(n))\Big\}-N_d\log\{\mathcal{M}_d\}.
\end{eqnarray}

Finally, taking the logarithm of (\ref{likelihood}), we have the \emph{\textbf{ninth term}} of (\ref{fe}) given by
\begin{eqnarray}\label{9}
&&\int_{{\boldsymbol\mu},\beta,\mathbf{x}_d} Q({\boldsymbol\mu})Q(\beta)Q(\mathbf{x}_d) \log p(\mathbf{y|}{\boldsymbol\mu},\beta,\mathbf{x}_d)
    d{\boldsymbol\mu}d\beta d\mathbf{x}_d \nonumber\\
    &=& -N\log\pi + N\mathbb{E}_{\beta}\{\log\beta \}    -\mathbb{E}_{\beta}\{\beta\}\Big[\mathbf{y}^H\mathbf{y}- 2\mathrm{\mathrm{Re}}\{\mathbf{y}^H\mathbf{G}\mathbf{[E}_p\mathbf{x}_p + \mathbf{E}_d\mathbf{\tilde x}_d]\mathbb{E}_{\mu}\{{\boldsymbol\mu}\}\}\nonumber\\
    &&+ \mathrm{Tr}\big\{\mathbf{G}^H[\mathbf{E}_p\mathbf{x}_p + \mathbf{E}_d\mathbf{\tilde x}_d]\mathbf{G}[\mathbf{E}_p\mathbf{x}_p+\mathbf{E}_d\mathbf{\tilde x}_d]\mathbb{E}_{\mu}\{{\boldsymbol\mu}{\boldsymbol\mu}^H\}\big\}\Big]\nonumber\\
    &=& -N\log\pi + N\left(\Psi(\tilde c) -\log(\tilde d)\right)-\left(\frac{\tilde c}{\tilde d}\right)\Big[\mathbf{y}^H\mathbf{y}- 2\mathrm{\mathrm{Re}}\left\{\mathbf{y}^H\mathbf{G}\mathbf{[E}_p\mathbf{x}_p + \mathbf{E}_d\mathbf{\tilde x}_d]\mathbf{\tilde m}_{\mu}\right\} \nonumber\\
    &&+ \mathrm{Tr}\left\{\mathbf{G}^H[\mathbf{E}_p+\mathbf{x}_p + \mathbf{E}_d\mathbf{\tilde x}_d]\mathbf{G}[\mathbf{E}_p\mathbf{x}_p+\mathbf{E}_d\mathbf{\tilde x}_d]\left(\mathbf{\tilde m}_{\mu}\mathbf{\tilde m}_{\mu}^H + \mathbf{\tilde \Sigma}_{\mu}\right)\right\}\Big].
\end{eqnarray}

\section{Derivation of Updating Functions}\label{B}
\begin{flushleft}
{\underline{{Updating $(\mathbf{\tilde m}_{\mu}, \mathbf{\tilde \Sigma}_{\mu})$ given $\tilde a_j, \tilde b_j, \tilde c, \tilde d$ and $\mathbf{\tilde x}_d$}}}
\end{flushleft}
Focusing on the terms in  $\mathbb{F}$ related to $\mathbf{\tilde \Sigma}_{\mu}$, we have
\begin{eqnarray}
    \frac{\partial\mathbb{F}}{\partial \mathbf{\tilde \Sigma}_{\mu}}
    &=& \frac{\partial}{\partial\mathbf{\tilde \Sigma}_{\mu}}\left\{-\log\det(\mathbf{\tilde \Sigma}_{\mu}) + \mathrm{Tr}\{\mathrm{diag}\{[\frac{\tilde a_1}{\tilde b_1},\ldots, \frac{\tilde a_M}{\tilde b_M}]\} \mathbf{\tilde \Sigma}_{\mu}\} \nonumber+ \frac{\tilde c}{\tilde d}[\mathbf{G}^H[\mathbf{\tilde x}]\mathbf{G}[\mathbf{\tilde x}] \mathbf{\tilde \Sigma}_{\mu}]\right\}\nonumber\\
    &=&
    -\mathbf{\tilde \Sigma}_{\mu}^{-1} + \mathrm{diag}\left\{\left[\frac{\tilde a_1}{\tilde b_1},\ldots, \frac{\tilde a_M}{\tilde b_M}\right]\right\} + \frac{\tilde c}{\tilde d}\mathbf{G}^H[\mathbf{\tilde x}]\mathbf{G}[\mathbf{\tilde x}].\label{d1}
\end{eqnarray}
Setting (\ref{d1}) to zero leads to (\ref{o1}).
On the other hand,
\begin{eqnarray}
  \frac{\partial \mathbb{F}}{\partial\mathbf{\tilde m}_{\mu}} &= &
  \frac{\partial}{\partial\mathbf{\tilde m}_{\mu}}\bigg\{
 \mathrm{Tr}\left\{\mathrm{diag}\left\{\left[\frac{\tilde a_1}{\tilde b_1},\ldots,
 \frac{\tilde a_M}{\tilde b_M}\right]\right\}
 \mathbf{\tilde m}_{\mu}\mathbf{\tilde m}_{\mu}^H\right\}
  + \frac{\tilde c}{\tilde d}\Big[-2\mathrm{\mathrm{Re}}\left\{\mathbf{y}^H\mathbf{G}[\mathbf{\tilde x}]\mathbf{\tilde m}_{\mu}\right\} \nonumber\\
  &&+\mathbf{G}^H[\mathbf{\tilde x}]\mathbf{G}[\mathbf{\tilde x}] \mathbf{\tilde m}_{\mu}\mathbf{\tilde m}_{\mu}^H\Big]\bigg\}\nonumber\\
  &=&\frac{\partial}{\partial\mathbf{\tilde m}_{\mu}}\left\{
  \mathbf{\tilde m}_{\mu}^H\left[\mathrm{diag}\left\{\left[\frac{\tilde a_1}{\tilde b_1},\ldots,
  \frac{\tilde a_M}{\tilde b_M}\right]\right\}+ \frac{\tilde c}{\tilde d}\mathbf{G}^H
  \left[\mathbf{\tilde x}\right]\mathbf{G}[\mathbf{\tilde x}]\right]\mathbf{\tilde m}_{\mu} -\frac{\tilde c}{\tilde d}
  \mathbf{\tilde m}_{\mu}^H\mathbf{G}^H[\mathbf{\tilde x}]\mathbf{y}\right\}\nonumber\\
  &= &\left[\mathrm{diag}\left\{\left[\frac{\tilde a_1}{\tilde b_1},\ldots,
  \frac{\tilde a_M}{\tilde b_M}\right]\right\} + \frac{\tilde c}{\tilde d}\mathbf{G}^H
  [\mathbf{\tilde x}]\mathbf{G}[\mathbf{\tilde x}]\right]\mathbf{\tilde m}_{\mu}
  -\frac{\tilde c}{\tilde d}\mathbf{G}^H[\mathbf{\tilde x}]\mathbf{y}.\label{d2}
\end{eqnarray}
Setting (\ref{d2}) to zero leads to (\ref{o2}).
\begin{flushleft}
{\underline{{Updating $(\tilde a_j, \tilde b_j)$ given $\mathbf{\tilde m}_{\mu}, \mathbf{\tilde \Sigma}_{\mu}, \tilde c, \tilde d$ and $\mathbf{\tilde x}_d$}}}
\end{flushleft}
Gathering the terms in $\mathbb{F}$ that are related to $\tilde a_j$, we have
\begin{eqnarray}
\frac{\partial\mathbb{F}}{\partial\tilde a_j}
&=&\frac{\partial}{\partial\tilde a_j}\bigg\{ \mathrm{Tr}\left\{\mathrm{diag}
\left\{\left[\frac{\tilde a_1}{\tilde b_1},\ldots, \frac{\tilde a_M}{\tilde b_M}\right]\right\}
\left(\mathbf{\tilde m}_{\mu}\mathbf{\tilde m}_{\mu}^H + \mathbf{\tilde \Sigma}_{\mu}\right)\right\}-\Psi(\tilde a_j)\nonumber \\
  && + \left[\tilde a_j\log\tilde b_j + (\tilde a_j-1)\left[\Psi(\tilde a_j) -\log\tilde b_j\right] -\tilde a_j -\log\Gamma(\tilde a_j)\right]\nonumber\\
  && -\left[ (a_j-1)\left[\Psi(\tilde a_j) -\log\tilde b_j\right] -b_j\tilde a_j/\tilde b_j\right]\bigg\}\nonumber\\
  &=& \frac{1}{\tilde b_j} \left[\left|[\mathbf{\tilde m}_{\mu}]_{j}\right|^2 + \mathbf{[\tilde \Sigma}_{\mu}]_{j,j}\right] - \Psi'(\tilde a_j) + \log\tilde b_j + (\tilde a_j-1)\Psi'(\tilde a_j)\nonumber\\
  && +\Psi(\tilde a_j) - \log\tilde b_j -1 -\Psi(\tilde a_j) -(a_j-1)\Psi'(\tilde a_j)+ \frac{b_j}{\tilde b_j}\nonumber\\
  &=& (\tilde a_j-a_j-1)\Psi'(\tilde a_j) -1 + \frac{1}{\tilde b_j}\left[\left|[\mathbf{\tilde m}_{\mu}]_{j}\right|^2 + \mathbf{[\tilde \Sigma}_{\mu}]_{j,j}+ b_j\right].\label{d3}
\end{eqnarray}
Similarly,
\begin{eqnarray}
\frac{\partial\mathbb{F}}{\partial\tilde b_j} &= &\frac{\partial}{\partial\tilde b_j}\bigg\{ \mathrm{Tr}\left\{\mathrm{diag}\left\{\left[\frac{\tilde a_1}{\tilde b_1},\ldots, \frac{\tilde a_M}{\tilde b_M}\right]\right\} \left(\mathbf{\tilde m}_{\mu}\mathbf{\tilde m}_{\mu}^H + \mathbf{\tilde \Sigma}_{\mu}\right)\right\} +\log\tilde b_j \nonumber\\
  && + \left[\tilde a_j\log\tilde b_j - (\tilde a_j-1)\log\tilde b_j\right]  -\left[(a_j-1)(-\log\tilde b_j) -b_j\tilde a_j/\tilde b_j \right]\bigg\}\nonumber\\
  &=& -\frac{\tilde a_j}{\tilde b_j^2}\left[\left|[\mathbf{\tilde m}_{\mu}]_{j}\right|^2 + \mathbf{[\tilde \Sigma}_{\mu}]_{j,j}\right] + \frac{1}{\tilde b_j} + \frac{\tilde a_j}{\tilde b_j} -\frac{\tilde a_j -1}{\tilde b_j} +\frac{a_j -1}{\tilde b_j} -\frac{b_j\tilde a_j}{\tilde b_j^2}\nonumber\\
  &=& \frac{a_j +1}{\tilde b_j} - \frac{\tilde a_j}{\tilde b_j^2}\left[\left|[\mathbf{\tilde m}_{\mu}]_{j}\right|^2 + \mathbf{[\tilde \Sigma}_{\mu}]_{j,j}+ b_j\right].\label{d4}
\end{eqnarray}
Setting both (\ref{d3}) and (\ref{d4}) to zero and solving the simultaneous equations, we obtain (\ref{o3}) and (\ref{o4}).
\begin{flushleft}
{\underline{{Updating $(\tilde c, \tilde d)$ given $\mathbf{\tilde m}_{\mu}, \mathbf{\tilde \Sigma}_{\mu}, \tilde a_j, \tilde b_j$ and $\mathbf{\tilde x}_d$}}}
\end{flushleft}
Following the procedure in updating other parameters, we compute
\begin{eqnarray}\label{d5}
\frac{\partial\mathbb{F}}{\partial\tilde c} &=
& \frac{\partial}{\partial\tilde c}\bigg\{\left(\tilde c -1\right)\Psi(\tilde c) -\tilde c -\log\Gamma(\tilde c) -(c-1)\Psi(\tilde c)+ \frac{\tilde c d}{\tilde d}- N\Psi(\tilde c)\nonumber\\
&&   +\frac{\tilde c}{\tilde d}\Big[\mathbf{y}^H\mathbf{y} - 2\mathrm{\mathrm{Re}}\left\{\mathbf{y}^H\mathbf{G}\mathbf{[\tilde x]\tilde m}_{\mu}\right\}+ \mathrm{Tr}\left\{\mathbf{G}^H[\mathbf{\tilde x}]\mathbf{G}[\mathbf{\tilde x}]\left(\mathbf{\tilde m}_{\mu}\mathbf{\tilde m}_{\mu}^H + \mathbf{\tilde \Sigma}_{\mu}\right)\right\}\Big]\bigg\}\nonumber\\
&= & (\tilde c-1)\Psi'(\tilde c) + \Psi(\tilde c) - 1 - \Psi(\tilde c) -(c-1)\Psi'(\tilde c)+ \frac{d}{\tilde d} - N\Psi'(\tilde c) \nonumber\\
&&  +\frac{1}{\tilde d}\Big[\mathbf{y}^H\mathbf{y} - 2\mathrm{\mathrm{Re}}\left\{\mathbf{y}^H\mathbf{G}\mathbf{[\tilde x]\tilde m}_{\mu}\right\} + \mathrm{Tr}\left\{\mathbf{G}^H[\mathbf{\tilde x}]\mathbf{G}[\mathbf{\tilde x}](\mathbf{\tilde m}_{\mu}\mathbf{\tilde m}_{\mu}^H + \mathbf{\tilde \Sigma}_{\mu})\right\}\Big]\nonumber\\
&= & (\tilde c-c-N)\Psi'(\tilde c) - 1 + \frac{1}{\tilde d}\Big[d + \mathbf{y}^H\mathbf{y} - 2\mathrm{\mathrm{Re}}\left\{\mathbf{y}^H\mathbf{G}\mathbf{[\tilde x]\tilde m}_{\mu}\right\}\nonumber\\
&& +\mathrm{Tr}\left\{\mathbf{G}^H[\mathbf{\tilde x}]\mathbf{G}[\mathbf{\tilde x}](\mathbf{\tilde m}_{\mu}\mathbf{\tilde m}_{\mu}^H + \mathbf{\tilde \Sigma}_{\mu})\right\}\Big],
\end{eqnarray}
and
\begin{eqnarray}\label{d6}
\frac{\partial\mathbb{F}}{\partial\tilde d} &=& \frac{\partial}{\partial\tilde d} \Big\{c\log\tilde d + \frac{\tilde cd}{\tilde d} + N\log\tilde d
+ \frac{\tilde c}{\tilde d}\Big[\mathbf{y}^H\mathbf{y} - 2\mathrm{\mathrm{Re}} \left\{\mathbf{y}^H\mathbf{G}\mathbf{[\tilde x]\tilde m}_{\mu} \right\}  \nonumber\\
&&  \vphantom{\frac{\tilde a}{\tilde b}}
+\mathrm{Tr} \left\{\mathbf{G}^H[\mathbf{\tilde x}]\mathbf{G}[\mathbf{\tilde x}] (\mathbf{\tilde m}_{\mu}\mathbf{\tilde m}_{\mu}^H + \mathbf{\tilde \Sigma}_{\mu})\right\}\Big] \big\} \nonumber\\
&=&  - \frac{\tilde c}{\tilde d^2}\big[ d+ \mathbf{y}^H\mathbf{y} - 2\mathrm{\mathrm{Re}}
\{\mathbf{y}^H\mathbf{G}\mathbf{[\tilde x]\tilde m}_{\mu}\}
+\frac{c+N}{\tilde d}+\mathrm{Tr}\{\mathbf{G}^H[\mathbf{\tilde x}]\mathbf{G}
[\mathbf{\tilde x}] (\mathbf{\tilde m}_{\mu}\mathbf{\tilde m}_{\mu}^H
+ \mathbf{\tilde \Sigma}_{\mu} )\}\big].
\end{eqnarray}
Setting both (\ref{d5}) and (\ref{d6}) to zero and solving the two simultaneous equations, we obtain (\ref{o5}) and (\ref{o6}).
\bibliographystyle{IEEEtran}
\bibliography{referencesmaria} 



\end{document}